\documentclass{jfm}
\usepackage{caption}
\usepackage{graphicx}
\usepackage{booktabs}
\usepackage{newtxtext}
\usepackage{newtxmath}
\usepackage{natbib}
\usepackage{subcaption}
\usepackage{hyperref}
\hypersetup{
    colorlinks=true,      % 启用链接颜色
  linkcolor=blue,       % 设置链接（包括表格引用）为蓝色
  citecolor=blue,       % 设置文献引用的颜色为蓝色
  urlcolor=blue         % 设置URL的颜色为蓝色
}

\newcommand{\RomanNumeralCaps}[1]
\title{Marangoni modulation of coupled Rayleigh–Taylor and Faraday instabilities in vertically oscillated liquid films}
\author{
Jun Gao\aff{1,2},
Senlin Zhu\aff{1}
\corresp{\email{slzhu@buaa.edu.cn}},
Luca Brandt\aff{3},
Jianjun Tao\aff{4},
Qingfei Fu\aff{1,2}
\corresp{\email{fuqingfei@buaa.edu.cn}},
Lijun Yang\aff{1}
}
\affiliation{
\aff{1}School of Astronautics, Beihang University, Beijing 100191, PR China \\
\aff{2}Aircraft and Propulsion Laboratory, Ningbo Institute of Technology, Beihang University, Ningbo 315800, PR China \\
\aff{3}Department of Environment, Land and Infrastructure Engineering, Politecnico di Torino, Corso Duca degli Abruzzi 24, Torino 10129, Italy  \\
\aff{4}Department of Mechanics and Engineering Science, College of Engineering, Peking University, Beijing 100871, PR China
}

\begin{document}
\maketitle

\begin{abstract}
We investigate the Marangoni modulation of coupled Rayleigh–Taylor and Faraday instabilities in a vertically oscillated Newtonian liquid film carrying insoluble surfactants. Linear stability analysis using Floquet theory reveals that an increasing Marangoni number ($\mathrm{Ma}$) selectively suppresses subharmonic modes, driving the system into a harmonic-dominated regime. The interfacial response is found to be highly frequency-dependent. At low forcing frequencies, increasing $\mathrm{Ma}$ causes adjacent harmonic tongues to merge into a novel surfactant mode that migrates towards long wavelengths, ultimately coalescing with the RTI branch and fragmenting the dynamically stable window. Conversely, at high frequencies, surfactants monotonically elevate the harmonic instability threshold, significantly widening the stable parameter space. To uncover the underlying mechanisms, a long-wave asymptotic analysis is performed, demonstrating that the critical forcing amplitude factorizes into a static capillary-gravity margin and a dynamic elasto-inertial modulation, yielding a scaling law for the critical mode balance. Finally, nonlinear simulations based on a rigorous weighted-residual reduced model are utilized to dissect the spatial work performed by individual forces, which shows that surfactants modulate stability through phase-controlled Marangoni transport. In the RTI regime, increasing $\mathrm{Ma}$ reverses the transport direction and drives fluid into the peaks, inducing a transition from stabilization to destabilization. In the Faraday instability (FI) regime, the response exhibits a strong frequency dependence, governed by Marangoni transport that redistributes fluid away from interfacial peaks at high frequencies but toward them at low frequencies, thereby suppressing or enhancing the instability accordingly.
\end{abstract}


{\bf MSC Codes }  {\it(Optional)} Please enter your MSC Codes here
\section{Introduction}
The Rayleigh–Taylor instability (RTI) is a canonical interfacial phenomenon that occurs when a high-density fluid is accelerated into a lower-density fluid \citep{Sharp1984}. This fundamental hydrodynamic phenomenon is pivotal in a vast range of physical scales and disciplines. In astrophysics, it governs the morphology and mixing dynamics of supernova remnants \citep{Cabot2006}. 
In engineering, and particularly in inertial confinement fusion (ICF), the RTI represents a primary obstacle to achieving ignition. During the deceleration phase of the capsule implosion, the growth of the ablative RTI can cause the shell to break up and induce severe mixing between the fuel and the ablator, degrading the final compression \citep{Betti2016, Zhou2025}. 
The development of RTI is also a key mechanism in the atomization of liquid propellants for aerospace and combustion applications, where the instability's growth on planar liquid sheets or coaxial jets is crucial for the primary breakup that leads to efficient droplet formation and combustion \citep{Qin2018, Yang2020}.
Such interfacial flow instabilities, including the complex nonlinear waves observed in sheared liquid films \citep{Hu2024}, are pervasive in various industrial processes.
Given its profound and often performance-limiting impact, a deep understanding of the mechanisms to control the RTI is of paramount scientific and practical importance.

One of the most effective methods for controlling interfacial stability is the application of time-periodic forcing, a condition inherent to many systems, from launch vehicles subject to engine vibrations to microfluidic devices driven by oscillating pressure fields.
The pioneering experimental and theoretical work of \citet{Wolf1969, Wolf1970} first demonstrated the remarkable phenomenon of dynamic stabilization, showing that a sufficiently strong vertical oscillation can completely suppress the long-wavelength RTI at an unstable interface with inverted density stratification. 
\citet{Troyon1971} conducted a stability analysis of viscous oscillating interfaces, indicating that viscosity and surface tension are necessary conditions for complete suppression of RTI.
This dynamic stabilization effect continues to inspire modern research, leading to striking demonstrations such as the levitation of dense liquids, which allows objects to float beneath them \citep{Apffel2020}.
Further research has extended these concepts to more complex scenarios, including the modulation of RTI in sedimenting suspensions \citep{Zhu2024,Zhu2026} and the enhanced suppression effects arising from the coupling effect of multi-frequency forcing \citep{Zhu2025}.
The concept has also found direct application in addressing the grand challenges outlined above, \citet{Zhao2024,Zhao2025} recently proposed using temporally modulated laser pulses to dynamically stabilize the ablative RTI in ICF targets, offering a novel path towards higher fusion yields. 

However, while high-frequency oscillations can stabilize the long-wavelength gravity-driven RTI, they can simultaneously excite short-wavelength interfacial waves through parametric resonance. This phenomenon, first observed by \citet{Faraday1831} and now known as the Faraday instability (FI), has since become a major field of fluid dynamics, attracting extensive theoretical and experimental study. Research has progressed from foundational linear stability analyses for viscous fluids \citep{Kumar1994} to explorations of superlattice wave patterns excited by multi-frequency forcing \citep{Kudrolli1998}, localized Faraday waves generated by heterogeneous forcing \citep{Urra2019},  and the behaviour of FI in non-Newtonian fluids \citep{Ignatius2025}. Consequently, in a system subject to both a constant gravitational  acceleration and time-periodic modulation,  the interfacial dynamics are governed by a complex competition between the long-wavelength RTI and the short-wavelength FI. Investigating this interplay for thin liquid films via the long-wave evolution equation, \citet{sterman2017rayleigh} revealed that the coupling between high-wavenumber FI modes and low-wavenumber RTI modes can arrest the catastrophic growth associated with RTI, resulting in a non-rupturing, dynamically saturated state. 
Building upon this, recent work by \citet{Zhu2025b} further elucidated the coupling mechanism. They found that the interface amplitude undergoes a characteristic multi-stage evolution, consisting of an initial growth, a subsequent decrease, and eventual saturation. This behaviour was explained through a local average thickness model, which revealed the governing mechanism to be a mutual interaction of induction and suppression between the FI and RTI modes.
These studies highlight that the presence of vibration does not simply suppress one instability but rather creates a new, complex system where the nonlinear competition between co-existing modes is the central dynamic.

The interfacial dynamics in most practical applications are further complicated by the presence of surface-active agents (surfactants), which adsorb to the interface and locally alter its properties. As elegantly summarized in a recent perspective by \citet{Lohse2023}, the physics of surfactant-laden interfaces is a rich and active field of modern fluid mechanics. Their relevance is particularly acute in applications such as advanced propulsion systems, where surfactants are essential for stabilizing colloidal suspensions of high-energy-density nanoparticles in liquid fuels \citep{chakraborty2020}. 
Furthermore, the dynamics of other complex fluid systems, such as surfactant-laden viscoelastic threads, are profoundly influenced by surface-active agents and their induced surface rheology \citep{Li2023}.
Surfactants introduce Marangoni stresses, which arise from gradients in surface tension and act to resist tangential motion at the interface. For the classical RTI, \citet{Stewart2013} experimentally showed that insoluble surfactants stabilize the interface in long soap films under adverse acceleration. Their analysis demonstrated that an increased Marangoni number (the ratio of Marangoni stress to viscous stress) reduces the maximum growth rate and dominant wavenumber of RTI, analogous to the effect of a vertical temperature gradient.
The influence of surfactants on the Faraday instability is  more intricate.
Seminal theoretical works by \citet{Kumar2002, Kumar2004} using Floquet analysis revealed that the critical wavenumber and amplitude for FI onset strongly depend on the phase shift between surfactant concentration and interface displacement. Remarkably, the critical amplitude exhibits a minimum with increasing Marangoni number, implying that surfactant-laden interfaces can be more susceptible to high-wavenumber modes than clean interfaces.
\citet{Suman2008} demonstrated that in inertialess regimes where clean Newtonian interfaces remain stable, sufficiently strong Marangoni flows can trigger instabilities that persist from long to finite wavelengths.
Nonlinear simulations by \citet{Panda2025} captured the emergence of exotic ridge and hill patterns driven by bidirectional Marangoni flows, representing qualitatively new topological features absent in surfactant-free FI.
In the context of oscillatory film flows, \citet{Gao2006} found that insoluble surfactants stabilize the flow by raising the critical Froude number and narrowing unstable frequency bandwidths. \citet{Wang2023} extended this to viscoelastic films, revealing that surfactants decrease unstable frequency ranges while viscoelasticity shifts stability boundaries and introduces both traveling-wave and standing-wave modes, with finite-wavelength instabilities dominating at high frequencies.

Despite extensive research on the effects of oscillatory forcing and surfactants on interfacial stability, these two mechanisms have been largely studied in isolation. Previous studies on oscillatory systems have focused primarily on clean interfaces, revealing the competition between RTI stabilization and FI excitation \citep{sterman2017rayleigh, Zhu2025}. In parallel, surfactant-related research has explored their influence on either the classical RTI under constant acceleration \citep{Stewart2013} or the Faraday instability in stable stratifications \citep{Kumar2002, Kumar2004, Panda2025}. However, the coupled dynamics arising when both mechanisms act simultaneously on a gravitationally unstable interface remain unexplored. This configuration introduces a three-way interaction between gravity-driven RTI, parametrically-driven FI, and Marangoni stresses induced by surfactant redistribution. Several fundamental questions emerge: At the linear level, how do Marangoni stresses modify the critical conditions for dynamic stabilization and the parametric instability thresholds? Will surfactants preferentially damp the long-wavelength RTI or the short-wavelength FI, thereby acting as a modal selection mechanism? Could oscillatory surfactant redistribution excite new coupled instability modes? At the nonlinear level, how will Marangoni flows interact with the RTI-FI competition and alter the dynamically saturated states identified in clean systems? The present work addresses these questions through linear stability analysis and numerical simulations of a surfactant-laden interface under combined constant and oscillatory acceleration.

The paper is organized as follows. In Sec. \autoref{sec:formulation}, we present the governing equations and boundary conditions, followed by the nondimensionalization and linearization. In Sec. \autoref{Linear stability}, the linear stability of the system is investigated using Floquet theory, focusing on the effects of Marangoni stresses on instability thresholds, mode selection, and stability-boundary topology. In Sec. \autoref{Asymptotic long-wave analysis}, a long-wave asymptotic analysis is performed to reveal the mechanisms underlying the surfactant-induced modal transitions and  derive analytical scaling for the critical conditions. In Sec. \autoref{nonlinear revolution}, we examine the nonlinear evolution and dynamically saturated states of the interface, and further clarify the frequency-dependent role of surfactants through force--work analysis. Finally, the main conclusions and physical implications of the present study are summarized in Sec. \autoref{Conclusion}.

\begin{figure}
  \centerline{\includegraphics[width=0.8\linewidth]{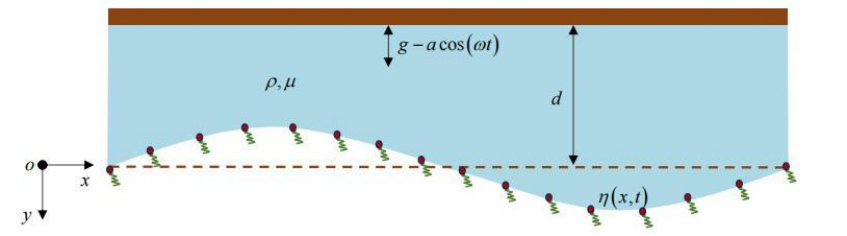}}
  \caption{\raggedright Schematic of subplate thin film flow under the combined effects of surfactants and vertical vibration.
 }
  \label{fig1}
\end{figure}

%Senlin
\section{Mathematical Formulation}
\label{sec:formulation}

\subsection{Governing equations}
Consider a horizontal layer of incompressible Newtonian fluid of density $\rho$ and dynamic viscosity $\mu$, with a mean thickness $d$, as shown in \autoref{fig1}. The fluid is bounded by a rigid plate at $y = -d$ and a free surface at $y = \eta(x, t)$. The system configuration is gravitationally unstable, representing a heavy liquid layer overlying a passive gas of negligible density and viscosity. A Cartesian coordinate system $(x, y)$ is attached to the oscillating plate, with the origin set at the equilibrium position of the free surface and the $y$-axis pointing towards the gas phase (parallel to the direction of gravity). The system is subjected to a vertical oscillatory acceleration $\mathbf{G}(t) = (g - a \cos \omega t)\mathbf{e}_y$, where $g$ is the gravitational acceleration, and $a$ and $\omega$ denote the amplitude and angular frequency of the oscillation, respectively.

The fluid motion is governed by the continuity and Navier-Stokes equations:
\begin{equation}
	\nabla \cdot \mathbf{u} = 0,
	\label{eqs2.1}
\end{equation}
\begin{equation}
	\rho \left[ \mathbf{u}_t + (\mathbf{u} \cdot \nabla)\mathbf{u} \right]
	= -\nabla p + \mu \nabla^2 \mathbf{u} + \rho \mathbf{G}(t).
	\label{eqs2.2}
\end{equation}
At the rigid plate ($y=-d$), the fluid satisfies the no-slip condition $\mathbf{u}=0$. At the deformable interface $y = \eta(x,t)$, the kinematic boundary condition is given by:
\begin{equation}
	\eta_{t} + (\mathbf{u} \cdot \nabla)\eta = v.
	\label{eqs2.3}
\end{equation}
And the stress balance is given by:
\begin{equation}
	 \mathbf{T} \cdot \mathbf{n} = \sigma \kappa \mathbf{n} + \nabla_{s} \sigma,
	\label{eqs2.4}
\end{equation}
where $\mathbf{T} = (p_{g} - p)\mathbf{I} + \mu \left[ \nabla \mathbf{u} + (\nabla \mathbf{u})^{T} \right]$ represents the total stress tensor incorporating the ambient gas pressure $p_g$, $\mathbf{n}$ is the outward unit normal vector, $\kappa = -\nabla_s \cdot \mathbf{n}$ is the mean curvature, and $\nabla_{s} = (\mathbf{I} - \mathbf{n}\mathbf{n}) \cdot \nabla$ is the surface gradient operator.
The evolution of the surfactant concentration $\Gamma$ on the interface is governed by the convection-diffusion equation:
\begin{equation}
	\Gamma_{t} + u \Gamma_{x} + \Gamma \left( \nabla_{s} \cdot \mathbf{u} \right)
	= D_{s} \nabla_{s}^{2} \Gamma,
	\label{eqs2.5}
\end{equation}
where $D_s$ is the surface diffusivity. The system is closed by a linear equation of state relating surface tension to surfactant concentration:
\begin{equation}
	\sigma(\Gamma) = \sigma_0 - E (\Gamma - \Gamma_0),
	\label{eqs2.6}
\end{equation}
where $\sigma_0$ is the surface tension at the reference concentration $\Gamma_0$, and $E = \frac{RT \Gamma_{\infty}}{\Gamma_{\infty} - \Gamma_{0}}$ is the Gibbs elasticity, with $\Gamma_{\infty}$, $R$, and $T$ representing the maximum surfactant concentration that the free surface can sustain, the universal gas constant, and the temperature, respectively.
The initial uniform concentration of the surfactant is set to $0.1\Gamma_{\infty}$, and the basic solution under the undisturbed state is  as follows:
\begin{equation}
	\mathrm{\bar{u}}=0,\bar{p}=\rho G(t) \mathbf{e}_y,
	\label{eqs2.7}
\end{equation}

\begin{table}
	\centering	  
    \begin{tabular}{llll}      		
		\textbf{Parameter} & \textbf{Description} & \textbf{Value} & \textbf{Unit} \\ 
		$\nu$       & Kinematic viscosity                     & $0.9$                & $\mathrm{mm^2/s}$ \\
		$\sigma_0$  & Surface tension (clean interface)       & $72$                 & $\mathrm{g/s^2}$ \\
		$\rho$      & Density                                & $9.972\times10^{-4}$ & $\mathrm{g/mm^3}$ \\
		$g$         & Gravitational acceleration             & $9.81\times10^{3}$   & $\mathrm{mm/s^2}$ \\
		$T$         & Temperature                            & $298.15$             & $\mathrm{K}$ \\
		$R$         & Universal gas constant                 & $8.31\times10^{9}$   & $\mathrm{mm^2\cdot g/(K\cdot mol\cdot s^2)}$ \\
		$\Gamma_\infty$ & Maximum surfactant concentration   & $10^{-11}$  & $\mathrm{mol/mm^2}$ \\
		$D_s$       & Surface diffusivity           & $10^{-2}$   & $\mathrm{mm^2/s}$ \\
		$d$         & Film thickness                         & $0.1$                & $\mathrm{mm}$ \\		
	
    \end{tabular}
	\caption{Physical parameters used in this study \citep{sterman2017rayleigh,LiS2023}.}
	\label{tab1}
\end{table}

\subsection{Nondimensionalization and Linearization}

We render the governing equations dimensionless by scaling the variables with the following characteristic quantities: length $d$, velocity $u_0 = \nu/d$ (\cite{sterman2017rayleigh}), time $d/u_0 = d^2/\nu$, pressure $\rho u_0^2$, and concentration $\Gamma_0$. The dimensionless variables are defined as follows:
\begin{equation}
	\left\{
	\begin{aligned}
		&\tilde{u} = u/u_0 , \quad \tilde{v} = v/u_0 , \quad \tilde{x} = x/d , \quad \tilde{y} = y/d , \quad \tilde{\eta} = \eta/d, \quad \tilde{\omega} = \frac{\omega d^2}{\nu}, \\
		&\tilde{\sigma} = \sigma/\sigma_0 , \quad \tilde{\Gamma} = \Gamma/\Gamma_0 , \quad \tilde{t} = \frac{t \nu}{d^2}, \quad \tilde{p} = (p - p_g)\frac{d^2}{\rho \nu^2}.
	\end{aligned}
	\right.
	\label{eqs2.8}
\end{equation}
Hereafter, the tildes are omitted for brevity. We impose infinitesimal perturbations on the quiescent base state ($\bar{\mathbf{u}}=0, \bar{\eta}=0, \bar{\Gamma}=1$). The flow variables are decomposed as $\Phi = \bar{\Phi} + \Phi'$, where the prime denotes the perturbation quantity. Substituting this decomposition into the dimensionless governing equations yields the linearized continuity and momentum equations:
\begin{equation}
	\left\{
	\begin{aligned}
		&\nabla \cdot \mathbf{u}' = 0, \\
		&\mathrm{Re}\, \mathbf{u}'_t = -\nabla p' + \nabla^2 \mathbf{u}',
	\end{aligned}
	\right.
	\label{eqs2.9}
\end{equation}
where the Reynolds number is $\mathrm{Re} = u_0 d / \nu \equiv 1$. To decouple the velocity field from the pressure perturbation, we apply the curl-curl operator (equivalently, taking the Laplacian of the vertical component) to the momentum equation. This eliminates the pressure gradient, resulting in a fourth-order evolution equation for the vertical velocity $v'$:
\begin{equation}
	\left( \mathrm{Re}\,\partial_t - \nabla^{2} \right)\nabla^{2} v' = 0.
	\label{eqs2.10}
\end{equation}

The linearized boundary conditions are derived as follows. At the rigid plate ($y=-1$), the no-slip and no-penetration conditions imply:
\begin{equation}
	v' = 0, \quad v'_y = 0.
	\label{eqs2.11}
\end{equation}
At the perturbed interface $y = \eta(x, t) \approx 0$, the linearized kinematic boundary condition relates the interface velocity to the fluid vertical velocity:
\begin{equation}
	\eta'_{t} = v'.
	\label{eqs2.12}
\end{equation}

For the stress balance, we define the horizontal gradient operator as $\nabla_H = \partial_x \mathbf{e}_x$ (in the two-dimensional context). The tangential stress boundary condition is obtained by taking the surface divergence of the stress tensor. Utilizing the continuity equation ($\partial_x u' + \partial_y v' = 0$), this condition simplifies to a balance between the viscous shear stress and the Marangoni stress:
\begin{equation}
	\left( \nabla_{H}^{2} - \partial_{yy} \right) v' = -\mathrm{Ma} \nabla_{H}^{2} \Gamma', \quad \text{at } y=0,
	\label{eqs2.13}
\end{equation}
where $\mathrm{Ma} = E\Gamma_{0}/(\mu u_{0})$ is the Marangoni number.

The derivation of the normal stress boundary condition requires the elimination of the pressure term. First, projecting the stress balance onto the normal vector and expanding the pressure to the first order yields:
\begin{equation}
	- \left[ p' + \left( B - A\cos(\omega t) \right) \eta' \right] + 2 v'_y 
	= \frac{1}{\mathrm{Ca}} \nabla_H^2 \eta',
	\label{eqs2.14}
\end{equation}
where $B = g d^2 / (\nu u_0)$ and $A= a d^2 / (\nu u_0)$ represent the dimensionless gravitational and oscillatory acceleration amplitudes, respectively, and $\mathrm{Ca} = \mu u_0 / \sigma_0$ is the Capillary number. To eliminate $p'$, we take the horizontal divergence of the linearized momentum equation (\ref{eqs2.9}) and utilize the continuity equation, which provides an explicit relation for the pressure Laplacian:
\begin{equation}
	\nabla_H^2 p' = \left( \mathrm{Re} \, \partial_t - \nabla^2 \right) v'_y.
	\label{eqs2.15}
\end{equation}
Applying the operator $\nabla_{H}^2$ to equation (\ref{eqs2.14}) and substituting equation (\ref{eqs2.15}) allows us to remove the pressure term completely, yielding the final form of the normal stress boundary condition:
\begin{equation}
	- \left[ \left( \mathrm{Re} \partial_t - \nabla^2 \right) v'_y 
	+ \left( B - A \cos(\Omega t) \right) \nabla_H^2 \eta' \right] 
	+ 2 \nabla_H^2 v'_y 
	= \frac{1}{\mathrm{Ca}} \nabla_H^4 \eta'.
	\label{eqs2.16}
\end{equation}

Finally, the linearized surfactant transport equation, governed by convection and surface diffusion, is given by:
\begin{equation}
	\Gamma'_t = v'_y + \frac{1}{\mathrm{Pe}} \nabla_H^2 \Gamma',
	\label{eqs2.17}
\end{equation}
where $\mathrm{Pe} = u_0 d / D_s$ is the surface Péclet number. The system is thus closed by equations (\ref{eqs2.10}), (\ref{eqs2.11}), (\ref{eqs2.12}), (\ref{eqs2.13}), (\ref{eqs2.16}), and (\ref{eqs2.17}). The dimensionless parameters used in this study are fixed as $\mathrm{Re}=1$, $B=12.11$, $\mathrm{Pe}=90$, and $\mathrm{Ca}=1.12 \times 10^{-4}$, while the Marangoni number $\mathrm{Ma}$ varies in the range $[0, 340]$.

\section {Linear stability analysis at arbitrary wavenumbers}
\label{Linear stability}
\subsection {Perturbation expansion and solution}
Given the periodicity of the base flow induced by the oscillatory acceleration, the stability of the system is governed by linear equations with time-periodic coefficients. In accordance with Floquet theory, we seek solutions for the interface deformation $\eta'$, surfactant concentration $\Gamma'$, and vertical velocity $v'$ in the following form:
\begin{equation}
	\left[ \eta'(x,t), \, \Gamma'(x,t), \, v'(x,y,t) \right] = e^{\mathrm{i}kx} \sum_{n=-\infty}^{\infty} e^{[s+\mathrm{i}(\alpha + n \omega)]t} \left[ \eta_n, \, \Gamma_n, \, v_n(y) \right] + \text{c.c.},
	\label{eqs3.1}
\end{equation}
where $k$ is the real wavenumber, $s$ is the real growth rate associated with the disturbances (typically $s=0$ for neutral stability boundaries), and $\alpha$ is the Floquet exponent which determines the periodicity of the response. The term $\mathrm{c.c.}$ denotes the complex conjugate. The integer $n$ represents the harmonic index of the expansion. Substituting the expansion (\ref{eqs3.1}) into the linearized bulk equation (\ref{eqs2.10}) yields a modified Orr--Sommerfeld equation for each temporal mode $n$:
\begin{equation}
	\left( D^2 - k^2 \right) \left( D^2 - q_n^2 \right) v_n(y) = 0,
	\label{eqs3.2}
\end{equation}
where $D \equiv \mathrm{d}/\mathrm{d}y$ and $q_n^2 = k^2 + \mathrm{Re}[s + \mathrm{i}(\alpha + n\omega)]$.
The general solution for the vertical velocity $v_n(y)$ is:
\begin{equation}
	v_n = C_{1n} e^{ky} + C_{2n} e^{-ky} + C_{3n} e^{Q_n y} + C_{4n} e^{-Q_n y}.
	\label{eqs3.3}
\end{equation}
where $\{C_{jn}\}$ are integration constants to be determined by the boundary conditions.

The coupling between the flow field, surfactant transport, and interfacial dynamics is established through the linearized boundary conditions at the wall ($y=-1$) and the interface ($y=0$). Substituting the ansatz (\ref{eqs3.1}) and solution (\ref{eqs3.3}) into equations (\ref{eqs2.11})--(\ref{eqs2.13}) and (\ref{eqs2.17}) yields a linear system relating the coefficients $\{C_{jn}\}$ to the interface amplitude $\eta_n$\citep{Kumar1994}.
Substituting the perturbation expansion (\ref{eqs3.1}) into the normal stress balance condition (\ref{eqs2.16}) yields:
\begin{equation}
	\sum_{n=-\infty}^{\infty} e^{(\gamma + \mathrm{i} n\omega)t} \left\{ 
	\left[ D^3 - \left( 3k^2 + \mathrm{Re} \gamma_n \right) D \right] v_n 
	+ \left[ B - \frac{k^4}{\mathrm{Ca}} - A \cos(\omega t) \right] \eta_n 
	\right\}  = 0.
	\label{eqs3.4}
\end{equation}
The parametric forcing term, $A \cos(\omega t)$, couples the $n$-th temporal mode with its neighbors $n \pm 1$, leading to the following rearranged recurrence relation\citep{Kumar2004}:
\begin{equation}
	\mathcal{S} \eta = A \mathcal{J} \boldsymbol{\eta},
	\label{eqs3.5}
\end{equation}
where $\boldsymbol{\eta} = [\dots, \eta_{-2}, \eta_{-1}, \eta_{0}, \eta_{1}, \eta_{2}, \dots]^\mathrm{T}$. Here, $\mathcal{S}$ contains the coefficients for the individual interface modes, and $\mathcal{J}$ represents the coupling between different temporal modes.
For numerical computation, the Fourier series is truncated at a finite order $N$, rendering $\mathcal{S}$ and $\mathcal{J}$ as square matrices of dimension $(2N+1) \times (2N+1)$.
In the present study, spectral convergence is achieved with $N=10$. We distinguish between two classes of instability modes based on the Floquet exponent $\alpha$: harmonic modes ($\alpha = 0$)  and subharmonic modes ($\alpha = \omega/2$). For the neutral stability boundaries ($s=0$), equation (\ref{eqs3.5}) constitutes a generalized eigenvalue problem for the critical forcing amplitude $A$ at a given wavenumber $k$, which is solved using the QZ algorithm. Conversely, to determine the growth rate $s$ for a fixed amplitude $A$, the determinant of the characteristic matrix is solved iteratively. 

While the Floquet analysis presented above is valid for arbitrary wavenumbers, a reduced-order model based on the weighted residuals method is derived in Appendix~\ref{appA} to facilitate theoretical analysis of the underlying mechanisms. The consistency between the original model and the reduced model is verified in Appendix~\ref{appB}, where excellent agreement is observed. The validity of the long-wave approximation requires the Womersley number, defined as $W = (\omega/2\pi) d^2 / \nu$, to be much smaller than unity \cite{sterman2017rayleigh}. Consequently, the frequency range investigated in this work is constrained to $f \in [10, 50]$ Hz. Unless otherwise stated, the solutions presented before \autoref{Asymptotic long-wave analysis} are obtained from the full model, whereas those presented thereafter are based on the reduced-order model. In the following section, we first examine the individual effects of surfactants and vertical oscillations on the RTI before exploring their coupled dynamics.

\subsection  {Separate effects of surfactant and vertical oscillation on RTI.}

\begin{figure}
    \centering
    % 第一行
    \begin{subfigure}[b]{0.45\textwidth}
        \centering      \includegraphics[width=\textwidth]{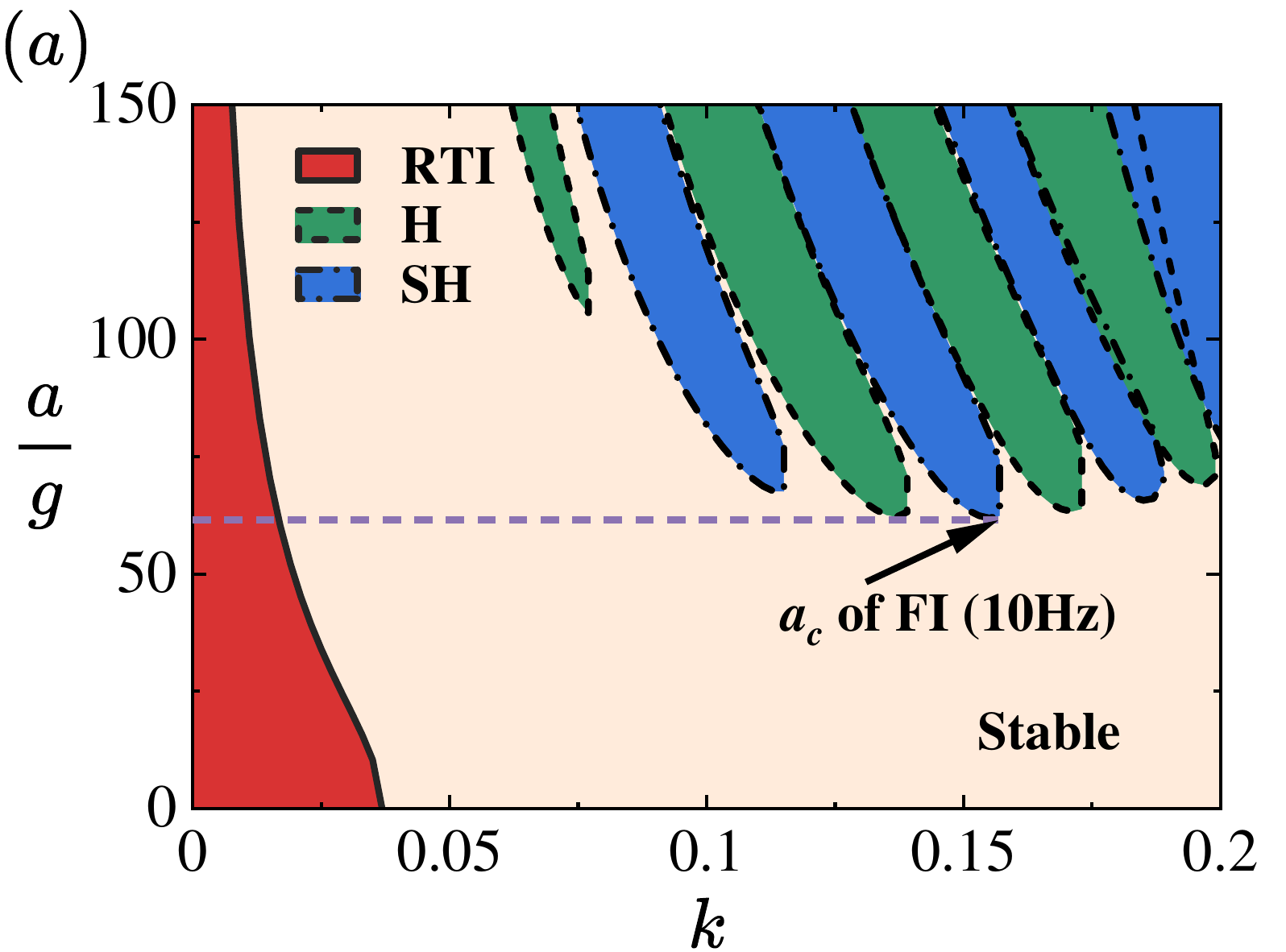}
        \label{fig2a}
    \end{subfigure}
    \hspace{0.02\textwidth}
    \begin{subfigure}[b]{0.45\textwidth}
        \centering        \includegraphics[width=\textwidth]{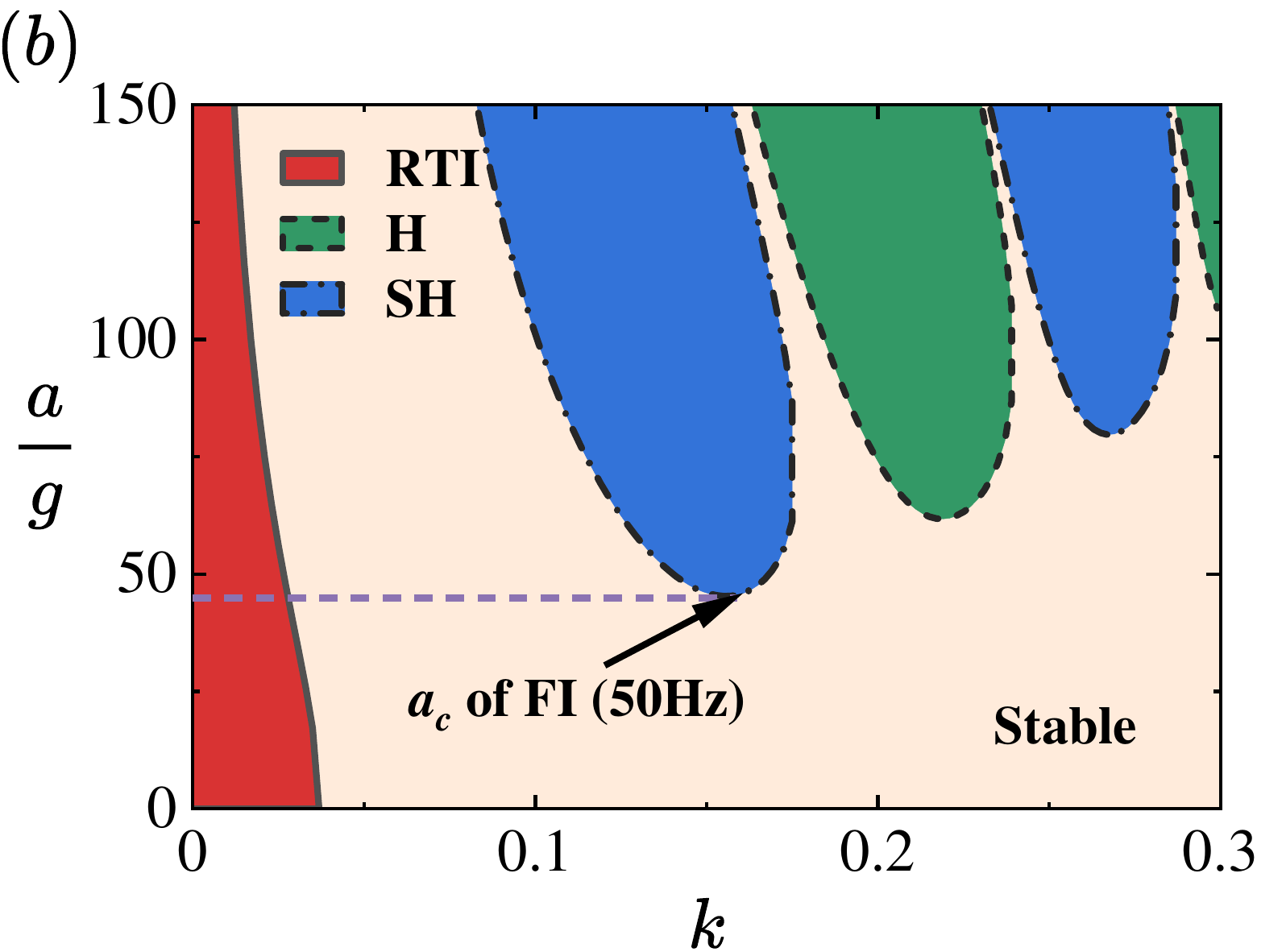}
        \label{fig2b}
    \end{subfigure}
    \vspace{0\textwidth} % 行间距
 % 第二行
    \begin{subfigure}[b]{0.45\textwidth}
        \centering       \includegraphics[width=\textwidth]{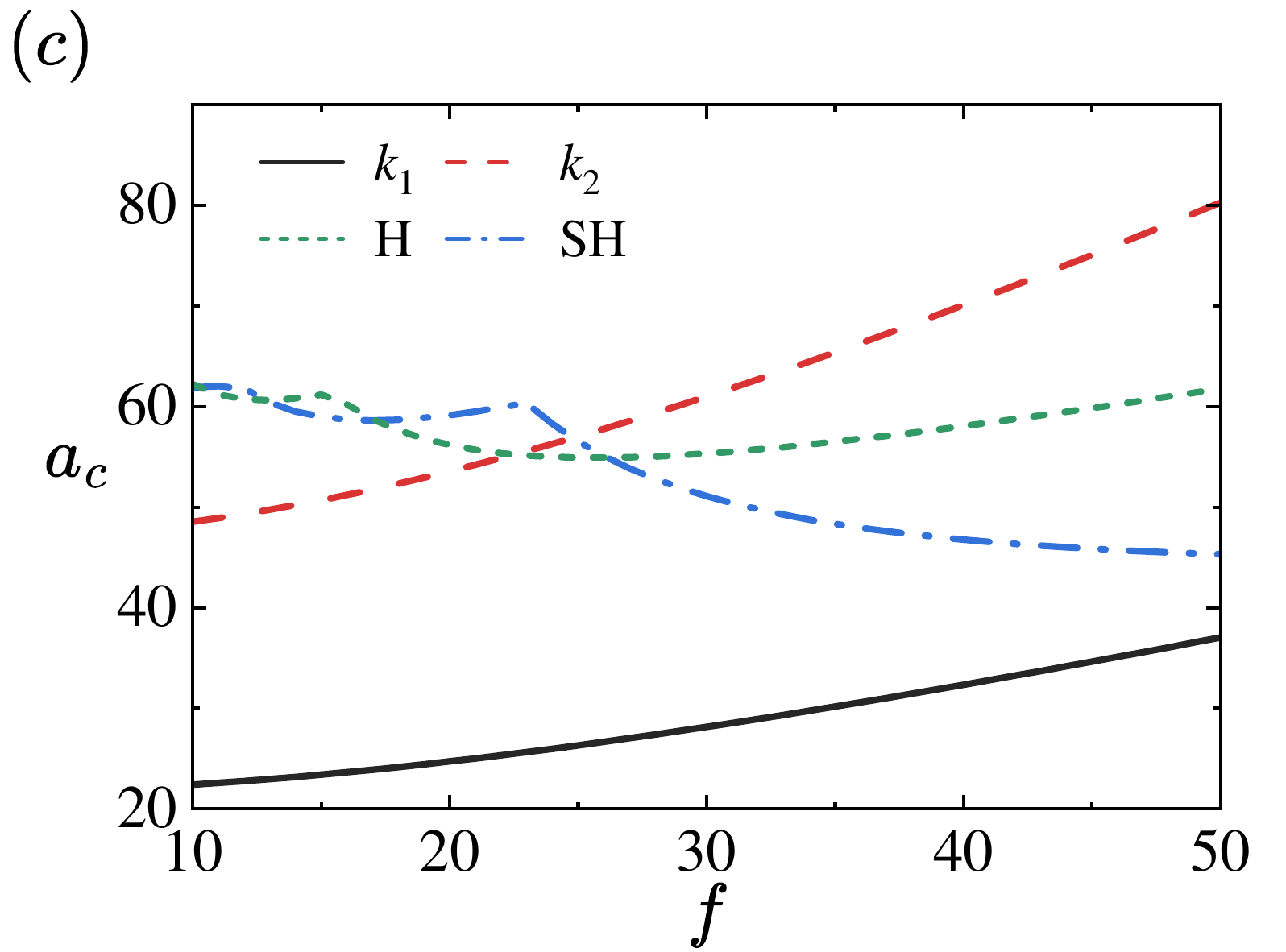}
        \label{fig2c}
    \end{subfigure}
    \hspace{0.02\textwidth}
    \begin{subfigure}[b]{0.45\textwidth}
        \centering       \includegraphics[width=\textwidth]{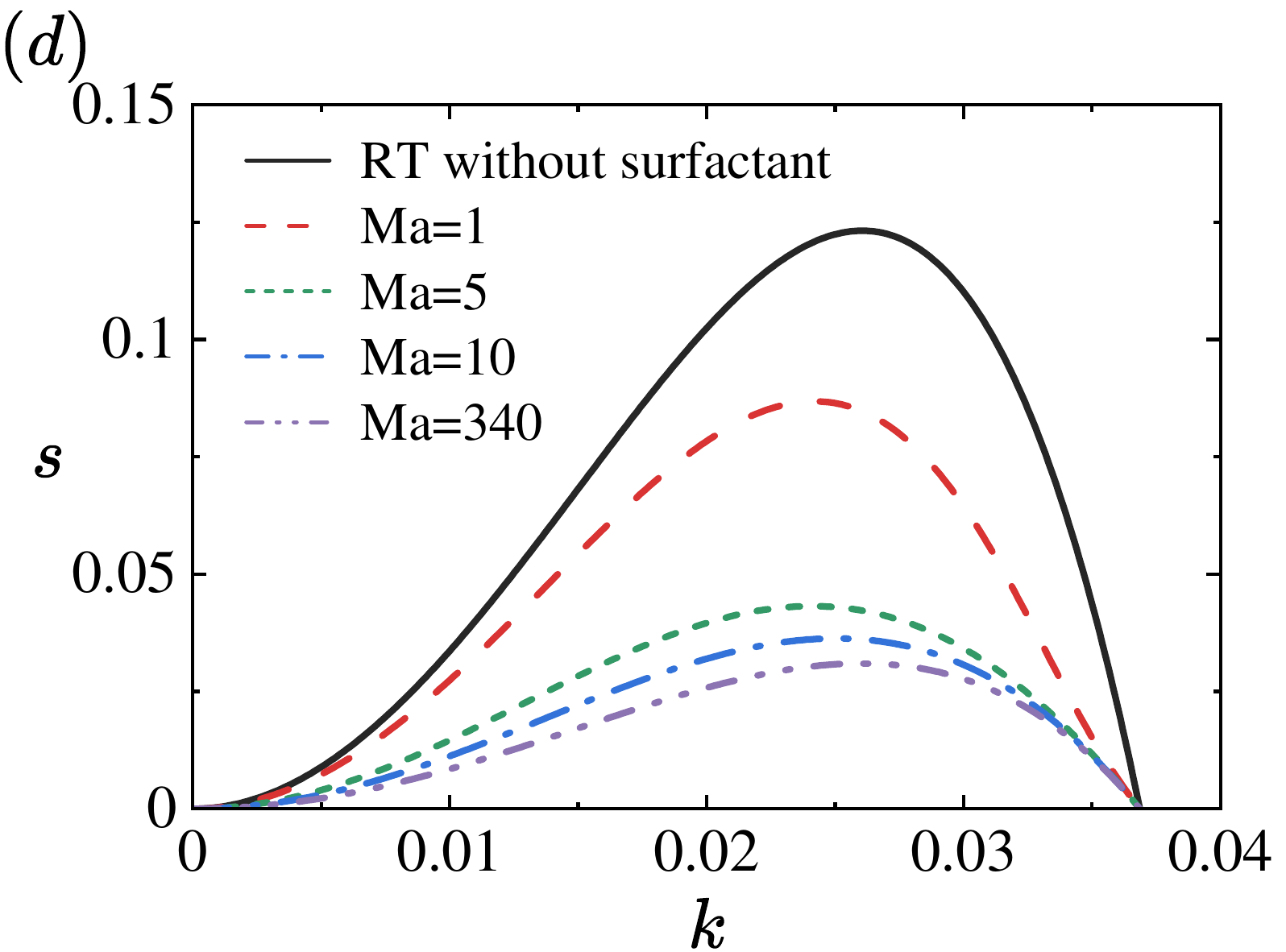}
        \label{fig2d}
    \end{subfigure}   
    \caption{ Separate effects of surfactants and vertical vibration on the stability of a liquid film beneath a plate. Neutral stability characteristics of a liquid film in the $a/g$–$k$ plane under $(a)$ low frequency $f=10$ and $(b)$ high frequency $f=50$. $(c)$ Critical amplitudes $a_c$ for RTI at wavenumbers $k_1 = 0.02$ and $k_2 = 0.03$, together with the variation of critical amplitudes for harmonic (H) and subharmonic (SH) modes with frequency $f$. $(d)$ Growth rate $s$–$k$ curves in the absence of vibration for different Marangoni numbers.}
    \label{fig2}
\end{figure}

The system considered here involves RTI, FI, and surfactant-induced Marangoni effects. As a reference, the effects of vertical vibration and surfactants on RTI are first examined separately. \autoref{fig2}($a-b$) present the neutral stability amplitude–wavenumber relations for RTI and FI ( including harmonic and subharmonic modes ) under vertical oscillations at 10\,Hz and 50\,Hz. With increasing oscillation amplitude, the liquid film can be dynamically stabilized against RTI. However, vertical vibration simultaneously introduces FI in the form of `tongue-shaped' instability regions. Consequently, effective suppression of RTI requires the oscillation amplitude to lie between the critical stability threshold of RTI and the onset threshold of FI. The `tongue-shaped' distributions in \autoref{fig2}($a-b$) differ markedly between the two frequencies. At 10\,Hz, the FI threshold is set by the second subharmonic and is higher than that at 50\,Hz, where the instability is governed by the first subharmonic mode. This behavior results from the competition between the viscous boundary layer thickness $\delta_v=\sqrt{2\nu/\omega}$ and the film thickness $d$ (\cite{Kumar1996}). 
When $\delta_v>d$ (low frequency), vibration-induced shear penetrates the entire film, leading to a globally viscous-dominated response, enhanced dissipation, and narrowed tongues, thereby elevating the FI threshold. 
In contrast, when $\delta_v<d$ (high frequency), dissipation is confined to a thin near-boundary layer, producing wider tongues and a lower critical amplitude.

\autoref{fig2}($c$) shows the variation of the critical amplitudes of FI and RT with frequency, where two representative RT wavenumbers, $k_1=0.02$ and $k_2=0.03$, are considered. Unlike the behaviour reported for thicker liquid films, in which the critical amplitude of FI typically increases with frequency, the present thin film system exhibits a different trend. At sufficiently high frequencies, the critical amplitude of the harmonic modes decrease initially and then increases with frequency, whereas that of the subharmonic modes decrease monotonically and remains lower than the harmonic counterpart. As a result, FI at high frequencies is controlled by the subharmonic modes, and its critical amplitude decreases with increasing frequency. A comparison of the RT critical amplitudes at the two selected wavenumbers further indicates that increasing the oscillation frequency suppresses RT instability only at $k=0.03$. Therefore, increasing the vibration frequency alone is insufficient for effective control of RTI in the present system.

\autoref{fig2}($d$) shows the dispersion relation between the dimensionless temporal growth rate $s$ and wavenumber $k$ under different Marangoni numbers $\mathrm{Ma}$. First, the $s$ of the subplate liquid film in the long-wave regime is greater than 0, which is due to the long-wave instability caused by RTI. As $\mathrm{Ma}$ increases, the temporal growth rate decreases, but it cannot be reduced to a value less than 0 in the long-wave regime. Meanwhile, the introduction of surfactant does not change the cutoff wavenumber of RTI.

Above all, vertical vibration can stabilize RTI only within a limited amplitude and frequency range, meanwhile, surfactants can reduce the growth rate of RTI but cannot eliminate the long-wave instability, indicating that neither mechanism alone provides effective control of RTI in the present system.

\begin{figure}
    \centering
    \begin{subfigure}[b]{0.45\textwidth}
        \centering
        \includegraphics[width=\textwidth]{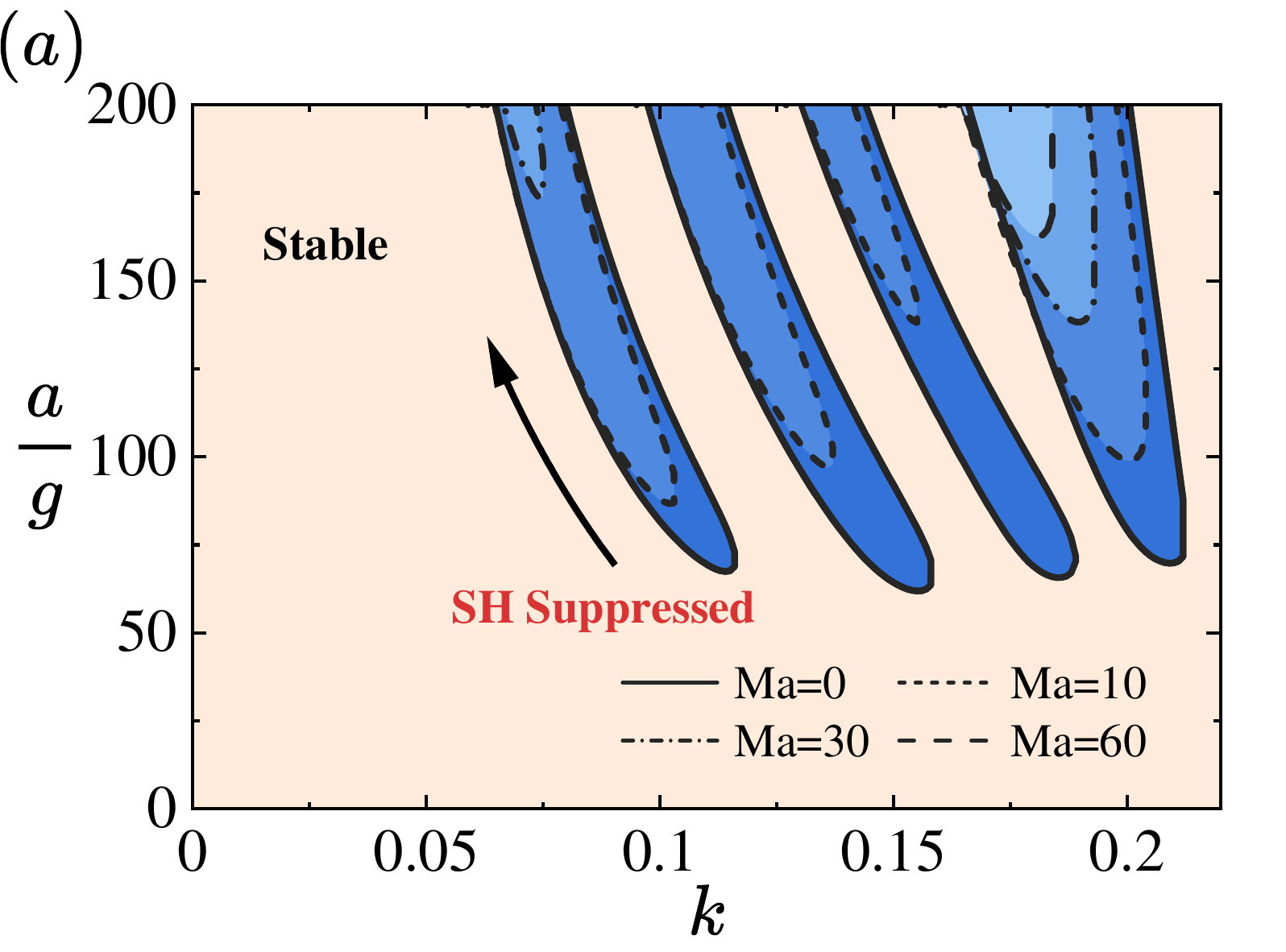}
       
    \end{subfigure}
    \hspace{0.02\textwidth}
    \begin{subfigure}[b]{0.45\textwidth}
        \centering
        \includegraphics[width=\textwidth]{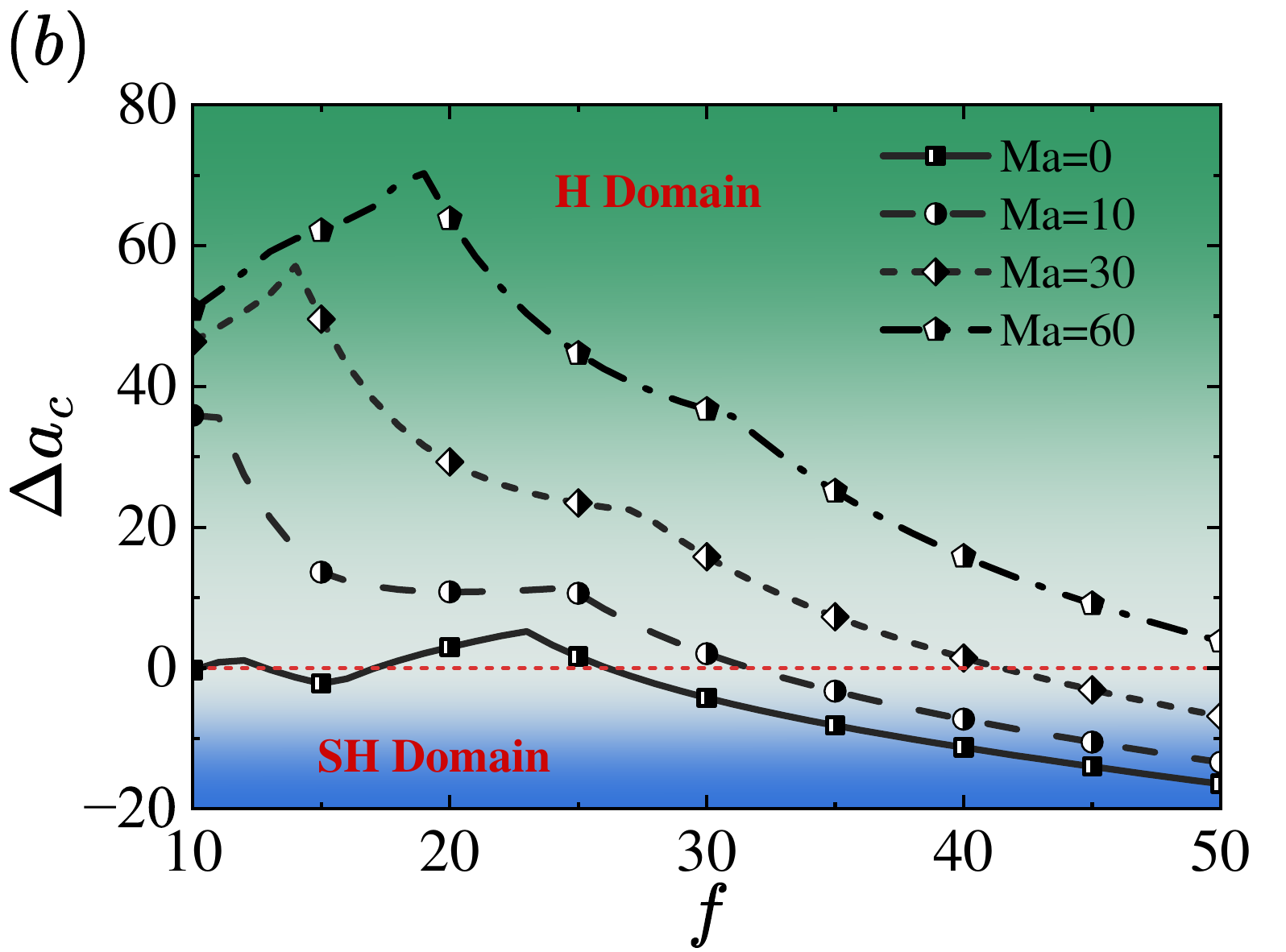}      
    \end{subfigure}
    \caption{Marangoni-induced suppression of subharmonic modes. $(a)$ Neutral stability curves of the subharmonic modes with varying $\mathrm{Ma}$.  
$(b)$ Variations of the critical amplitudes of harmonic and subharmonic modes with $\mathrm{Ma}$ at different frequencies. Here, $\Delta a_c = a_c^{\mathrm{SH}} - a_c^{\mathrm{H}}$, where
$a_c^{\mathrm{H}}$ and $a_c^{\mathrm{SH}}$ denote the critical amplitudes for the onset of harmonic and subharmonic modes, respectively.
}
    \label{fig3}
\end{figure}

\begin{figure}
    \centering
    % 第一行
    \begin{subfigure}[b]{0.45\textwidth}
        \centering
        \includegraphics[width=\textwidth]{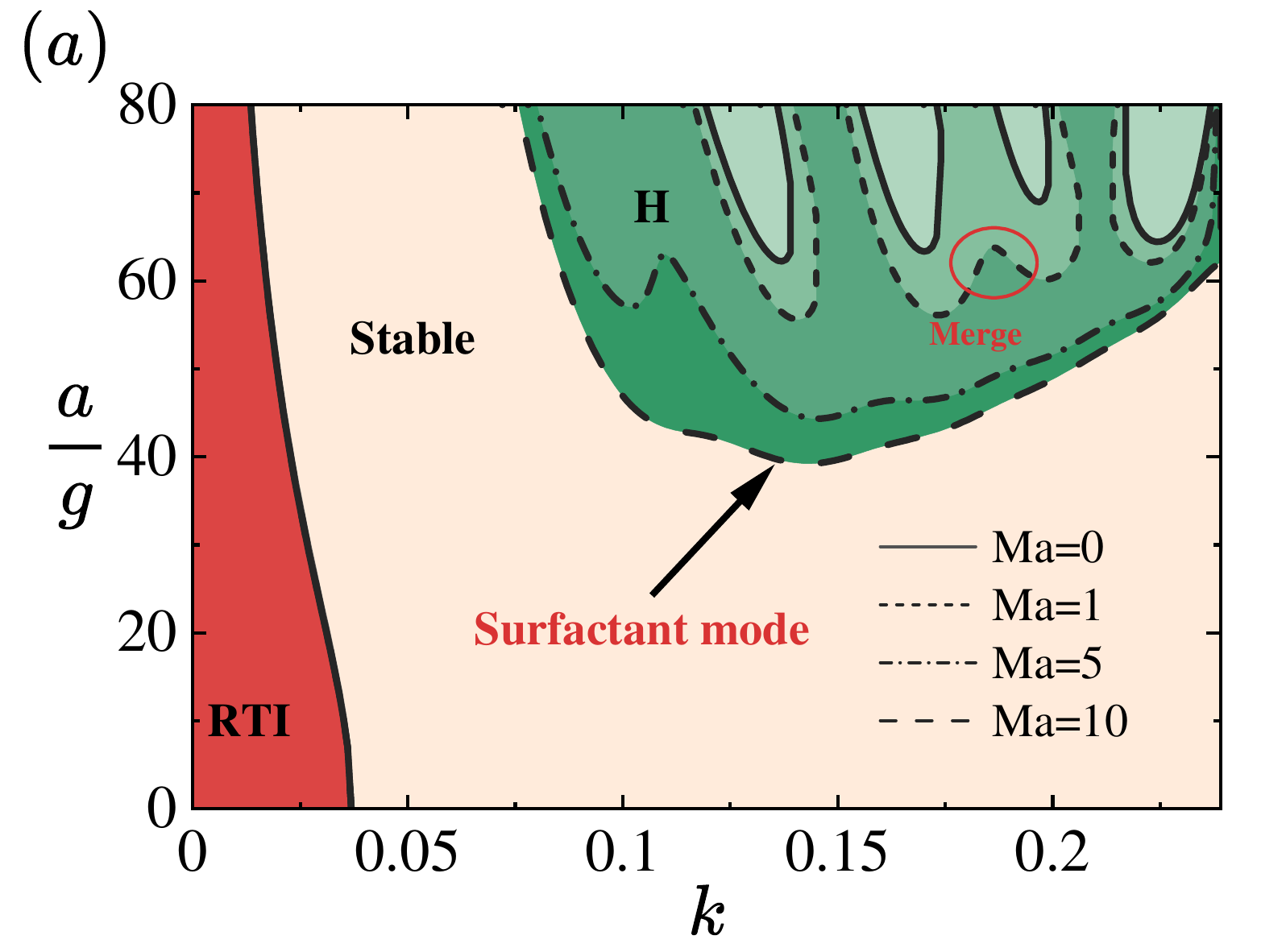}
       
    \end{subfigure}
    \hspace{0.02\textwidth}
    \begin{subfigure}[b]{0.45\textwidth}
        \centering
        \includegraphics[width=\textwidth]{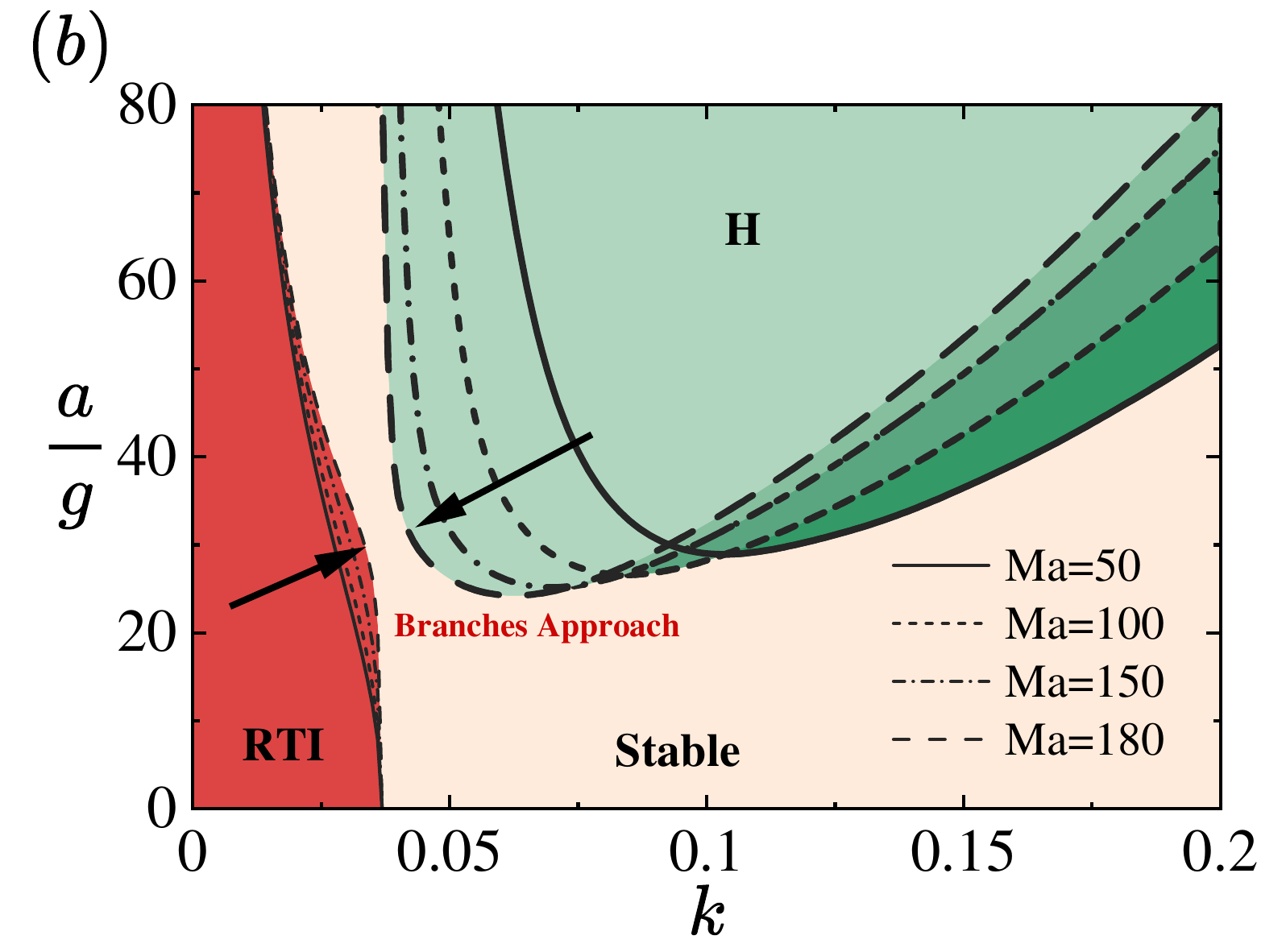}
        
    \end{subfigure}

    \vspace{0\textwidth} % 行间距

    % 第二行
    \begin{subfigure}[b]{0.45\textwidth}
        \centering
        \includegraphics[width=\textwidth]{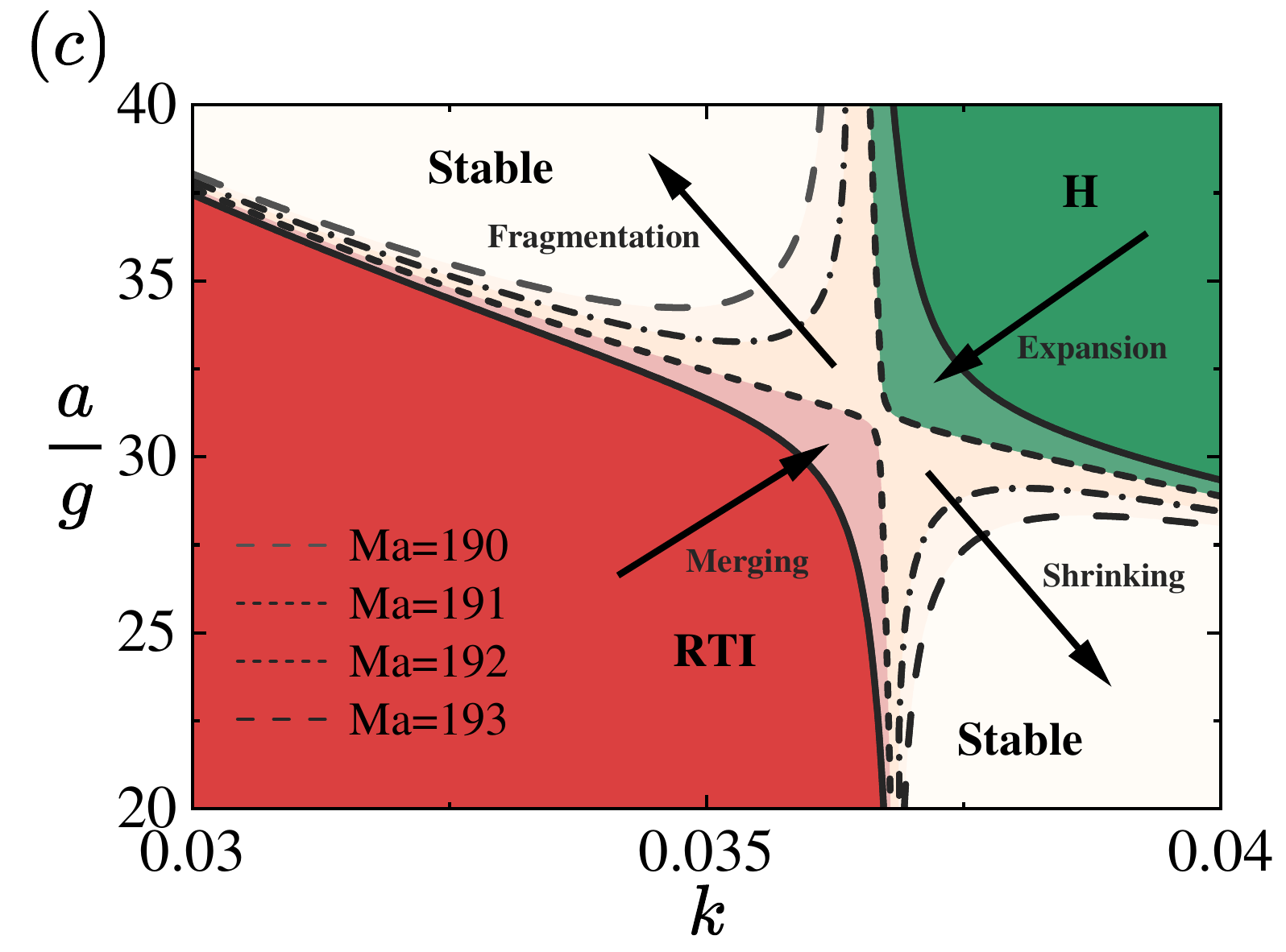}
        
    \end{subfigure}
    \hspace{0.02\textwidth}
   \begin{subfigure}[b]{0.45\textwidth}
       \centering
        \includegraphics[width=\textwidth]{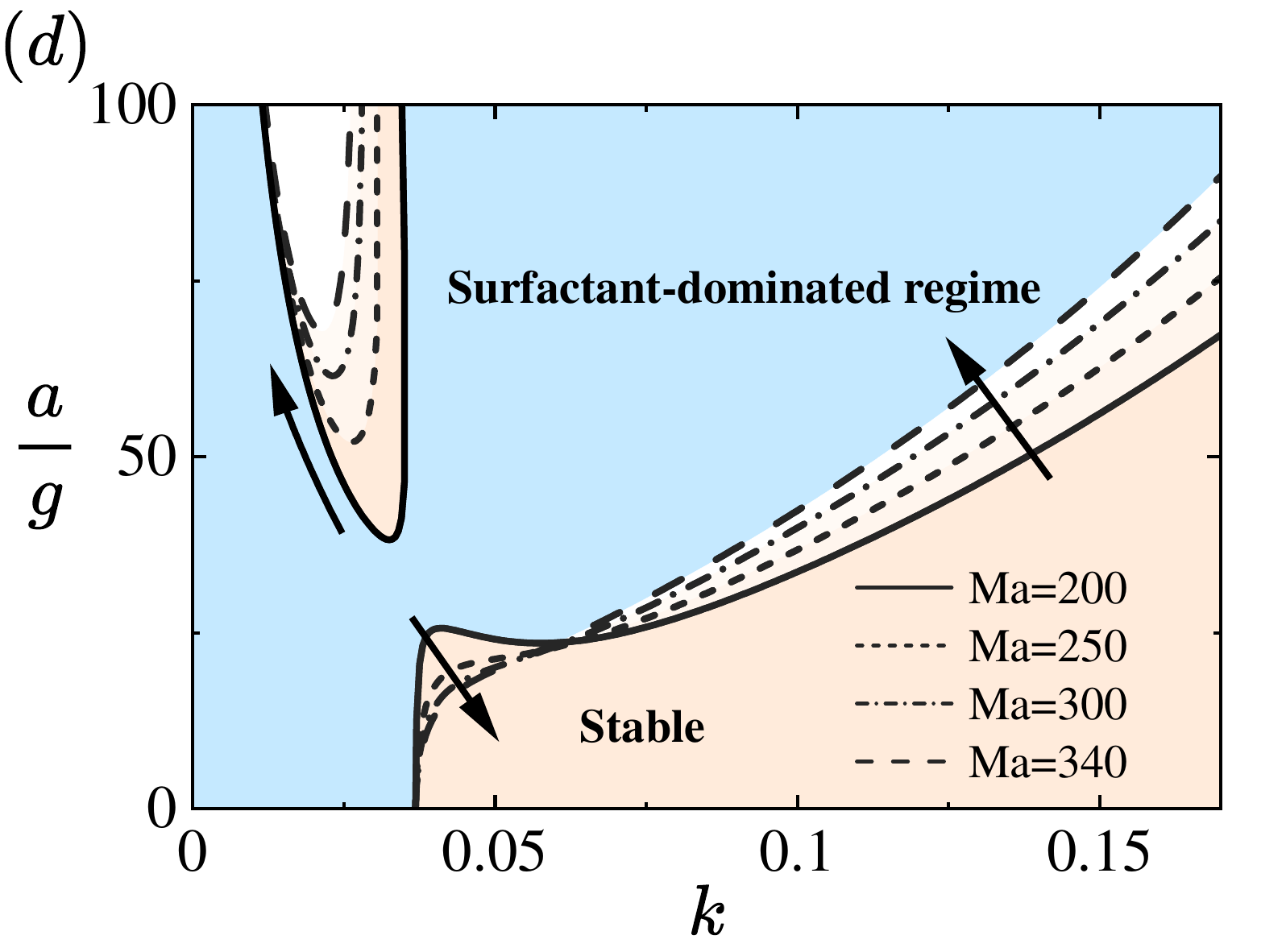}       
    \end{subfigure}
    \caption{  Marangoni-driven evolution and reorganization of instability modes at low frequency (10Hz)
($a$) Merging of neighbouring harmonic instability tongues with increasing $\mathrm{Ma}$, leading to the emergence of a surfactant-dominated mode.
($b$) Progressive approach between the RTI branch and the surfactant mode as $\mathrm{Ma}$ increases.
($c$) Merging of RTI and surfactant mode, accompanied by the expansion of unstable regions and the fragmentation of the stable domain.
($d$) Formation of a surfactant-dominated instability regime at large $\mathrm{Ma}$. }
    \label{fig4}
\end{figure}

\subsection {Marangoni effects on RTI and FI across frequency regimes} 

In this subsection, the effects of the Marangoni number $\mathrm{Ma}$ on subharmonic and harmonic modes are examined separately. \autoref{fig3}($a$) shows that the critical amplitude of the subharmonic mode increases monotonically with $\mathrm{Ma}$, indicating a pronounced stabilizing effect of surfactants on subharmonic responses. As further illustrated in \autoref{fig3}($b$), once $\mathrm{Ma}$ exceeds approximately 60, the instability threshold is governed by harmonic modes over the entire frequency range considered in this study. This behaviour reveals a mode-selective modulation induced by surfactants: with increasing $\mathrm{Ma}$, subharmonic modes are progressively suppressed, and the system transitions from a subharmonic-dominated regime to a harmonic-dominated regime. Owing to this clear separation in modal responses, the subsequent analysis focuses primarily on harmonic modes and RTI.

The reorganization of instability regions in the $a/g$--$k$ plane with increasing $\mathrm{Ma}$
is shown in \autoref{fig4}, revealing a qualitatively new behaviour in the low-frequency regime. At small $\mathrm{Ma}$, as shown in \autoref{fig4}($a$), the instability is dominated by multiple harmonic tongues that remain well separated, while the RTI branch exhibits little sensitivity to variations in $\mathrm{Ma}$, leaving a continuous stable region between the two unstable domains. As $\mathrm{Ma}$ increases, neighbouring harmonic tongues progressively merge, giving rise to a surfactant mode that is absent in clean systems. In this stage, the critical amplitude of RTI remains nearly unchanged, whereas the surfactant mode shifts towards lower wavenumbers and smaller amplitudes, as illustrated in \autoref{fig4}($b$), leading to a progressive approach between the two instability branches. \autoref{fig4}($c$) shows the subsequent merging process between the RTI mode and the surfactant mode. With a further increase in $\mathrm{Ma}$, the two instabilities first come into contact at a finite amplitude at the RTI cut-off wavenumber, resulting in the coalescence of unstable regions. This merging fragments the originally connected stable domain and causes it to shrink progressively as $\mathrm{Ma}$ increases. At sufficiently large $\mathrm{Ma}$, a surfactant-dominated instability regime emerges, as shown in \autoref{fig4}($d$), whereby the previously continuous stable region at low $\mathrm{Ma}$ becomes fragmented into two disconnected stability pockets. Although two stable subregions are still present in the $a/g$--$k$ plane and continue to evolve with increasing $\mathrm{Ma}$, effective stabilization of RTI is no longer achievable in this regime.

\begin{figure}
  \centerline{\includegraphics[width=0.6\linewidth]{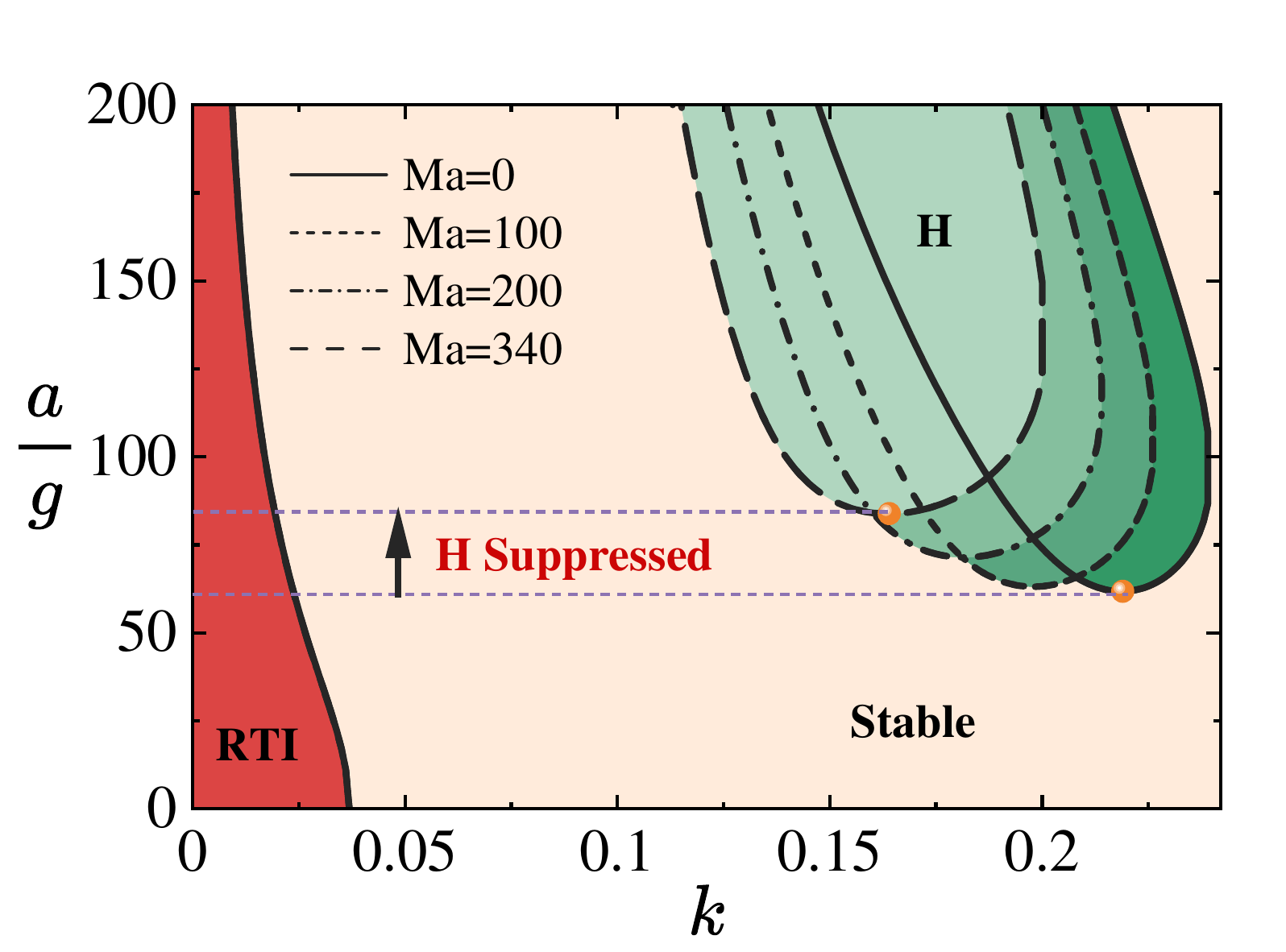}}
  \caption{ High-frequency (50Hz) modulation of instability regions by Marangoni effects.
 }
  \label{fig5}
\end{figure}

A markedly different behaviour is observed in the high-frequency regime, as shown in \autoref{fig5}. In this case, the critical amplitude of the harmonic instability increases significantly with increasing $\mathrm{Ma}$, leading to a substantial enlargement of the stable window between the Rayleigh–Taylor and Faraday instability thresholds. This result indicates that, unlike the low-frequency regime, the combined use of high-frequency forcing and surfactant addition allows effective suppression of RTI without triggering Faraday waves. 

In addition,from \autoref{fig4} and \autoref{fig5}, there exists a subtle but important feature that is easily overlooked: the introduction of surfactants significantly reduces the critical wavenumber of the harmonic mode, and this shift is nearly independent of the forcing frequency. By defining the wavenumber range associated with RTI as the long-wave regime and that of FI at high frequencies as the short-wave regime.  This comparison indicates that Marangoni effects modulate instability in fundamentally different ways in the long and short wave regimes, and the underlying mechanisms remain unclear. To elucidate the origin of this disparity and to identify the dominant balances controlling the instability, a long-wave asymptotic analysis is therefore employed.

\section {Asymptotic long-wave analysis of the reduced model} 
\label{Asymptotic long-wave analysis}
The numerical results presented in the previous section reveal that as the $\mathrm{Ma}$ increases, the Faraday instability becomes increasingly governed by harmonic modes. Crucially, when $\mathrm{Ma}$ is sufficiently large, the critical wavenumber of the first harmonic mode shifts towards the long-wave limit, eventually interacting with the Rayleigh–Taylor (RT) regime. To elucidate the physical mechanism underlying this surfactant-mediated coupling, we employ an asymptotic long-wave analysis based on the linearized reduced-order model derived in Appendix \ref{appB}.

To capture the essential physics with analytical clarity, we follow the approach utilized in previous stability analyses of complex fluids (e.g. \citet{Ignatius2025}) and truncate the Floquet expansion at order $N=1$ for the harmonic solution ($\alpha=0$), retaining only the zeroth ($n=0$) and first ($n=1$) temporal modes. Such a truncation strategy is physically justified as it isolates the interaction between the two dominant mechanisms identified in the numerical simulations: the RT mode ($n=0$), representing the non-oscillatory gravitational instability, and the surfactant-dominated harmonic mode ($n=1$), which captures the oscillatory Faraday response heavily modified by Marangoni stresses. Higher-order harmonics are neglected as they do not contribute to the leading-order coupling mechanism in the long-wave regime.

By invoking the harmonic reality condition ($\hat{\eta}_{-n} = \hat{\eta}_n^*$), the governing equations for the complex amplitudes $\hat{\eta}_0$ and $\hat{\eta}_1$ are assembled into a closed linear algebraic system. The condition for the existence of non-trivial solutions is that the determinant of the coefficient matrix must vanish. Noting that the zeroth-order mode represents a quasi-static deformation implies its coefficient is purely real ($S_0^i = 0$), the solvability condition takes the form:

\begin{equation}
\left|
\begin{matrix}
   S_{0}^{r} & 0 & -2A & 0  \\
   0 & S_{0}^{r} & 0 & 0  \\
   -A & 0 & S_{1}^{r} & -S_{1}^{i}  \\
   0 & -A & S_{1}^{i} & S_{1}^{r}  \\
\end{matrix}
\right| = 0.
\label{eq:determinant}
\end{equation}

Solving this determinant yields an explicit analytical expression for the neutral stability boundary $A(k)$:

\begin{equation}
A = \sqrt{ S_0^r \frac{ (S_1^r)^2 + (S_1^i)^2 }{ 2 S_1^r } }.
\label{eqs4.2}
\end{equation}

In this expression, $S_0^r$ represents the static stiffness of the interface, governed solely by the competition between gravity and capillarity. A negative value of $S_0^r$ corresponds to the classical Rayleigh–Taylor instability mechanism. The terms $S_1^r$ and $S_1^i$ constitute the real and imaginary parts of the dynamic coefficient for the first harmonic mode, encapsulating the effects of fluid inertia, viscous dissipation, and surfactant-induced Marangoni stresses. Their explicit forms, derived from the reduced-order model, are given by:

\begin{equation}
S_{0}^{r} = 2\left( B - \frac{1}{Ca} k^{2} \right),
\label{eqs4.3}
\end{equation}

\begin{equation}
S_{1}^{r} = \frac{12 \operatorname{Re}}{5 k^{2}} 
\left\{ 
\omega^{2} + \frac{5}{6 \operatorname{Re}} k^{2} \left( B - \frac{1}{Ca} k^{2} \right)
+ \frac{\left[ \frac{12(k^{2}-5)}{\operatorname{Re}} - \left( \frac{1}{Pe} + \frac{1}{4} \mathrm{Ma} \right) k^{2} \right] \omega^{2} k^{2} \mathrm{Ma}}{32 \left[ \omega^{2} + \left( \frac{1}{Pe} + \frac{1}{4} \mathrm{Ma} \right)^{2} k^{4} \right]} 
\right\},
\label{eqs4.4}
\end{equation}

\begin{equation}
S_{1}^{i} = \frac{12 \operatorname{Re}}{5 k^{2}} 
\left\{
- \left( \frac{5}{2 \operatorname{Re}} \omega + \frac{9}{2 \operatorname{Re}} k^{2} \omega \right)
+ \frac{\left[ \omega^{3} + \frac{12 (k^{2}-5)}{\operatorname{Re}} \left( \frac{1}{Pe} + \frac{1}{4} \mathrm{Ma} \right) k^{2} \omega \right] k^{2} \mathrm{Ma}}{32 \left[ \omega^{2} + \left( \frac{1}{Pe} + \frac{1}{4} \mathrm{Ma} \right)^{2} k^{4} \right]}
\right\}.
\label{eqs4.5}
\end{equation}

To extract the dominant physical mechanisms governing the interaction between the RT and harmonic modes, we perform a scaling analysis. In the frequency range considered in the present study, the dimensionless frequency $\omega$ is of order unity ($\omega \sim O(1)$). We focus on the long-wave regime in the vicinity of the RTI cutoff, where the relevant scaling relations for the control parameters are $B \sim O(k^{-1})$, $\mathrm{Pe} \sim O(k^{-1})$, and $\mathrm{Ca} \sim O(k^{-3})$. Under these conditions, a non-trivial dynamic coupling emerges specifically when the Marangoni number scales as $\mathrm{Ma} \sim O(k^{-2})$. This scaling implies that surfactant-induced stresses are of sufficient magnitude to compete with fluid inertia at the leading order.

Under these asymptotic conditions, the real and imaginary parts of the dynamic coefficient $S_1$ simplify to:

\begin{equation}
S_{1}^{r}=
\frac{12\operatorname{Re}}{5}\frac{{{\omega }^{2}}}{{{k}^{2}}}-\frac{9\mathrm{Ma}}{2} + o(\frac{1}{k^2}),
\label{eqs4.6}
\end{equation}

\begin{equation}
S_{1}^{i}=-6\frac{\omega }{{{k}^{2}}}
-\frac{9}{10}\frac{\mathrm{Ma}^2}{{k}^{2}}{\omega }+ o(\frac{1}{k^2}).
\label{eqs4.7}
\end{equation}

Substituting these asymptotic forms back into equation (\ref{eqs4.2}), the expression for the critical amplitude $A$ can be factorized into a product of a static term and a dynamic term. To highlight the physical distinctness of these contributions, we express the stability threshold as:
\begin{equation}
A(k) \approx \sqrt{ \mathcal{F}_S(k) \cdot \mathcal{F}_D(k, \omega, \mathrm{Ma}) }.
\label{eq:factorized_A}
\end{equation}

The first factor, $\mathcal{F}_S(k) = B - k^2/\mathrm{Ca}$, corresponds directly to the leading-order approximation of the static stiffness $S_0^r$ derived in the reduced model. A vanishing $\mathcal{F}_S$ recovers the classical static cutoff wavenumber, $k_{RT} = \sqrt{B \cdot \mathrm{Ca}}$. Crucially, this relationship implies that the range of wavenumbers susceptible to the static Rayleigh–Taylor instability is determined solely by the hydrostatic balance between gravity and capillarity, and remains invariant under the influence of vertical oscillations or surfactant addition.

The second factor, $\mathcal{F}_D$, represents the dynamic response of the harmonic mode. Based on the asymptotic structure of $S_1^r$ and $S_1^i$, this function takes the form:
\begin{equation}
\mathcal{F}_D \approx \left( \mathcal{I} - \mathcal{M} \right) + \frac{\mathcal{D}^2}{\mathcal{I} - \mathcal{M}}.
\label{eq:dynamic_func}
\end{equation}
Here, $\mathcal{I} = 12\mathrm{Re}\omega^2 / (5k^2)$ represents the effective inertial contribution, proportional to $\omega^2$ and inversely proportional to $k^2$. Notably, increasing frequency enhances the nominal inertial contribution but also shifts the dynamic critical wavenumber to higher values, where viscous dissipation is stronger. This increased dissipation offsets the inertial amplification, leading to a reduction in the effective inertia. $\mathcal{M} = 9\mathrm{Ma}/2$ denotes the Marangoni contribution originating from the tangential stress balance. The term $\mathcal{D} = 6\omega k^{-2} + 9\mathrm{Ma}^2 k^2 (10\omega)^{-1}$ corresponds to the magnitude of the residual imaginary component of the dynamic coefficient. Interestingly, $\mathcal{D}$ and $\mathcal{F}_D$ share a similar structure, from which the theoretical minimum of $\mathcal{F}_D$, $12\sqrt{15}/5 \mathrm{Ma}$, can be readily obtained, with the lower bound determined by the Marangoni effect.

% fig6
\begin{figure}
  \centering
  \begin{subfigure}{0.48\linewidth}
    \centering
    \includegraphics[width=\linewidth]{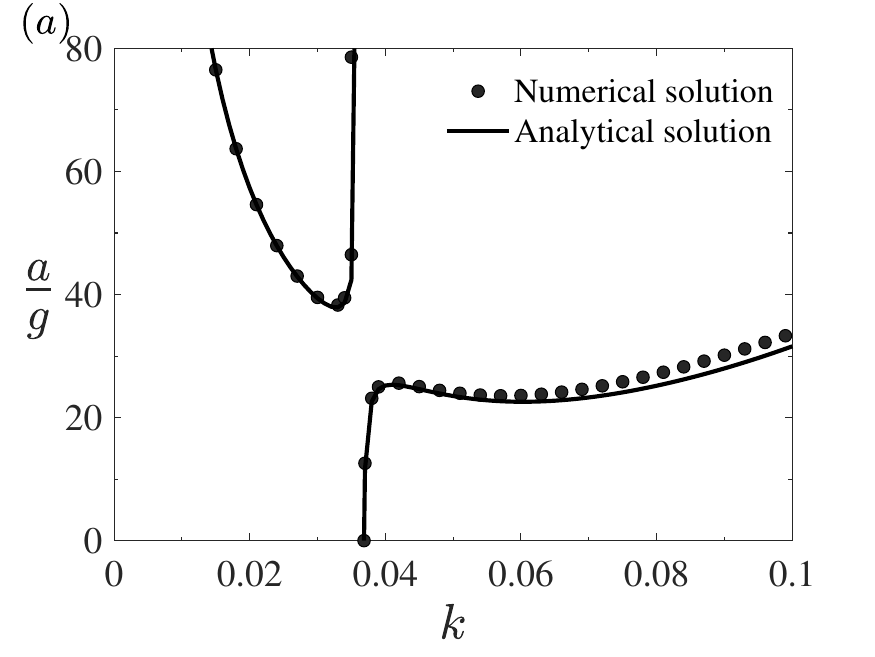}
    \label{fig6a}
  \end{subfigure}
  \hspace{0.02\textwidth}
  \begin{subfigure}{0.48\linewidth}
    \centering
    \includegraphics[width=\linewidth]{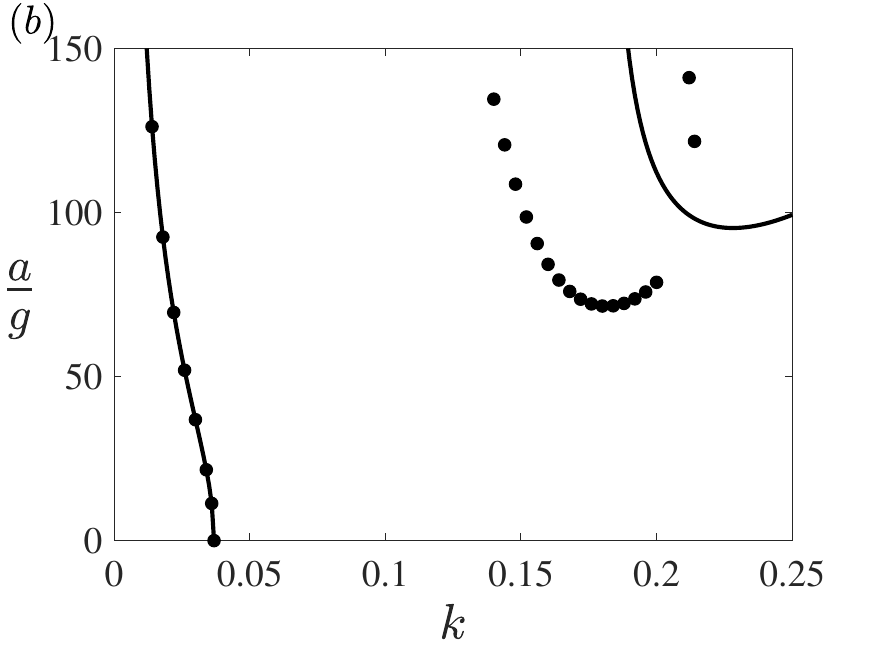}   
    \label{fig6b}
  \end{subfigure}
  \caption{Comparison between the long-wave analytical solution and numerical results, with the corresponding parameter values as follows: ($a$)$f=10\,\mathrm{Hz}, \mathrm{Ma}=200$. ($b$)$f=50\,\mathrm{Hz}, \mathrm{Ma}=200$.}
  \label{fig6}
\end{figure}

This factorization reveals that the system stability is determined by the interplay between the static margin $\mathcal{F}_S$ and the dynamic modulation $\mathcal{F}_D$. The structure of equation (\ref{eq:dynamic_func}) is particularly illuminating as it identifies the physical origin of the mode coalescence observed in the numerical simulations. The dynamic response exhibits a singularity when the effective inertia balances the Marangoni stress ($\mathcal{I} \approx \mathcal{M}$). This condition defines a critical dynamic balance, occurring at a specific crossover wavenumber $k_c$. In the regime $k < k_c$, the dynamics are dominated by effective inertia ($\mathcal{I} > \mathcal{M}$), whereas for $k > k_c$, the interface stiffens due to the dominance of Marangoni elasticity ($\mathcal{M} > \mathcal{I}$). The crossover point $k_c$ thus acts as a dynamic barrier, separating the classical inertia-driven Faraday instability from the surfactant-dominated regime. The quantitative location of this barrier is obtained by equating the effective inertial and Marangoni terms, yielding the scaling law for the critical wavenumber:
\begin{equation}
k_c = \sqrt{\frac{8 \mathrm{Re} \, \omega^2}{15 \mathrm{Ma}}}.
\label{eq:kc_scaling}
\end{equation}

This relation highlights the kinematic role of Marangoni stresses: increasing $\mathrm{Ma}$ monotonically shifts the surfactant-dominated mode towards longer wavelengths. \autoref{fig6} demonstrates excellent agreement between the analytical results derived from this factorization and the full numerical simulations, particularly in the low-frequency regime where the mode coupling is most pronounced.

The factorization $A \approx \sqrt{\mathcal{F}_S \mathcal{F}_D}$ provides a clear framework to interpret the distinct stability behaviors observed at different frequencies. In the high-frequency regime, the strong inertial contribution positions the dynamic critical wavenumber $k_c$ far above the static cutoff $k_{RT}$. In this decoupled limit ($k_c \gg k_{RT}$), the harmonic mode operates in a wavenumber range where the static factor $\mathcal{F}_S$ is large and insensitive to wavenumber variations. Consequently, the stability threshold is controlled primarily by the minimum value of the dynamic factor $\mathcal{F}_D$, which is $12\sqrt{15}/5 \mathrm{Ma}$. As $\mathrm{Ma}$ increases, the enhanced Marangoni stress raises the effective stiffness of the interface, increasing the minimum of $\mathcal{F}_D$. This mechanism explains the monotonic stabilization of the Faraday instability observed in \autoref{fig5}

A qualitatively different scenario emerges in the low-frequency regime, where the lower inertia places the initial $k_c$ closer to the RT cutoff. As $\mathrm{Ma}$ increases, $k_c$ shifts progressively towards $k_{RT}$. Although the local dynamic stiffness $\mathcal{F}_D$ technically increases with $\mathrm{Ma}$, the mode migrates into a region where the static stability margin $\mathcal{F}_S$ decays rapidly towards zero. Since the critical amplitude $A$ is proportional to the square root of the product $\mathcal{F}_S \mathcal{F}_D$, the reduction in $\mathcal{F}_S$ outweighs the increase in $\mathcal{F}_D$. This competition leads to a net decrease in the instability threshold, clarifying the physical origin of the surfactant-induced destabilization observed in \autoref{fig4}($a$).
% fig7
\begin{figure}
\centerline{\includegraphics[width=0.6\linewidth]{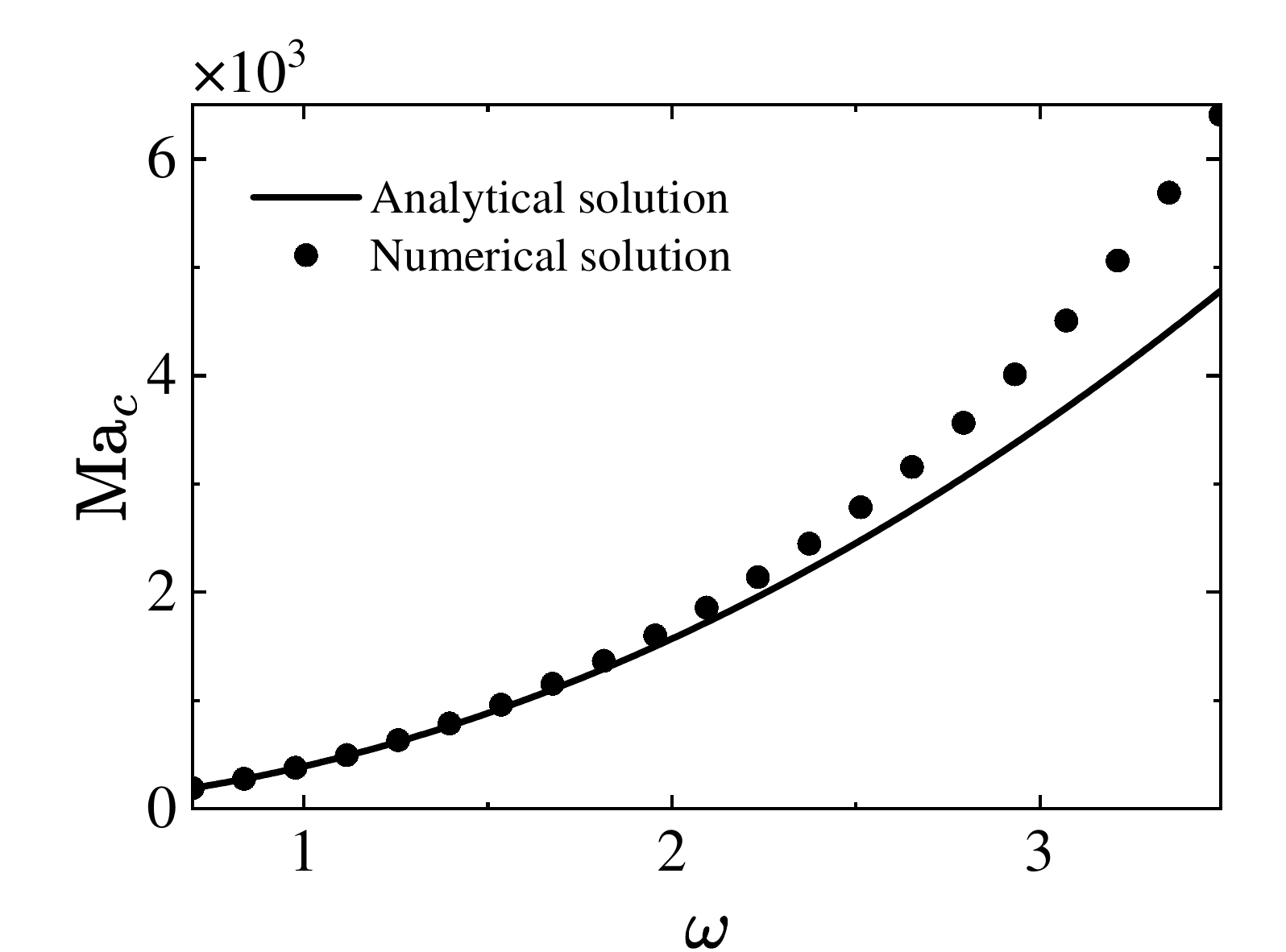}}
  \caption{\raggedright Prediction of the critical Marangoni number $\mathrm{Ma}_c$ for the fusion of the surfactant mode and RTI mode at different frequencies $\omega$.
 }
  \label{fig7}
\end{figure}
Furthermore, when $\mathrm{Ma}$ is sufficiently large, the dynamic singularity $k_c$ penetrates the RTI unstable band ($k < k_{RT}$). Mathematically, the divergence of $\mathcal{F}_D$ at $k_c$ intersects the RTI neutral curve, effectively severing the continuous unstable region. This corresponds to the physical process of mode coalescence and the fragmentation of the stable domain shown in \autoref{fig4}($b-c$). The onset of this transition is predicted by the condition $k_c = k_{RT}$, which yields a critical Marangoni number:
\begin{equation}
\mathrm{Ma}_c = \frac{8 \mathrm{Re} \, \omega^2}{15 B \cdot \mathrm{Ca}}.
\label{eq:Mac_prediction}
\end{equation}
The dependence of $\mathrm{Ma}_c$ on frequency is plotted in \autoref{fig7}, where good agreement between the analytical predictions and numerical results is obtained over the parameter range considered. This confirms that the topological transition of the stability boundaries is fundamentally driven by the interaction between the surfactant-induced Marangoni stress, dissipation-modulated effective inertia and the capillary-gravity waves.

\section{Nonlinear interfacial evolution and underlying mechanisms}
\label{nonlinear revolution}
In the preceding section, the influence of large $\mathrm{Ma}$ on the RTI and FI is analytically examined under low and high frequency conditions by employing a long wave asymptotic method. The competition between the increase in the critical amplitude induced by large $\mathrm{Ma}$ and the decrease in the critical amplitude associated with the reduction of the critical wavenumber is identified as the origin of the frequency dependent effects of surfactants. However, in the low-frequency regime at small $\mathrm{Ma}$, the RTI critical amplitude is nearly insensitive to $\mathrm{Ma}$, and the mechanism responsible for the reduction of the FI threshold remains unclear. In contrast, at high frequencies, the only moderate agreement between analytical and numerical results suggests that the mechanism by which $\mathrm{Ma}$ elevates the FI threshold remains to be clarified. These limitations motivate a transition to the nonlinear regime, where the underlying physical mechanisms can be examined in greater detail. In Appendix \ref{appB}, a reduced model is derived using a weighted residual approach as follows :
\begin{equation}
{{\eta }_{t}}+{{q}_{x}}=0,
\label{eqs5.1}
\end{equation}

\begin{equation}
{{\Gamma }_{t}} + \partial_x\left[ \frac{3 q \Gamma}{2 b} - \frac{\mathrm{Ma}}{4} b \Gamma \Gamma_{x} \right] = \frac{1}{Pe} \Gamma_{xx}.
\label{eqs5.2}
\end{equation}

\begin{equation}
\begin{aligned}
 & -\underbrace{\delta \mathrm{Re}\Big\{ \frac{2}{5} b^2 q_{t} + \frac{34 b}{35} q q_{x} - \frac{18}{35} q^2 \eta_{x} \Big\}}_{\text{inertial terms}} -\underbrace{\delta^3 \mathrm{Re} \mathrm{Ma}^2 \frac{b^4}{336} \Gamma_{x} \Big( \frac{5}{2} b \Gamma_{xx} + \frac{19}{5} \eta_{x} \Gamma_{x} \Big)}_{\text{inertial terms influenced by surfactant}} \\
 & +\underbrace{\delta^2 \mathrm{Re} \mathrm{Ma} \Big\{ \frac{b^3}{120} \big[ b \Gamma_{xt} + \frac{19}{7} q_{x} \Gamma_{x} \big] 
   + \frac{b^2}{56} q \big[ \Gamma_{x} \eta_{x} + \frac{3 b}{2} \Gamma_{xx} \big] \Big\}}_{\text{inertial term influenced by surfactant}} \\
 & = -\frac{b^3}{3} \underbrace{\Big( \delta B \eta_{x} - \delta A \cos(\omega t) \eta_{x} + \frac{\delta^3}{Ca} \eta_{xxx} \Big)}_{\text{gravity, oscillation, capillary terms}} \\
 & \quad + \underbrace{q + \delta^2 \Big\{ \frac{9}{5} b (q_{x} \eta_{x} - b q_{xx}) + \frac{4}{5} q (3 b \eta_{xx} - 2 \eta_{x}^2) \Big\}}_{\text{viscous term}} \\
 & \quad + \underbrace{\frac{1}{2} \delta b^2 \mathrm{Ma} \Gamma_{x} 
   + \delta^3 \mathrm{Ma} \frac{b^2}{5} \Big\{ \frac{b}{2} ( b \Gamma_{xxx} + 3 \eta_{x} \Gamma_{xx} ) + \frac{b}{3} \eta_{xx} \Gamma_{x} - 4 \eta_{x}^2 \Gamma_{x} \Big\}}_{\text{Marangoni stress terms}},
\end{aligned}
\label{eqs5.3}
\end{equation}
where $b=1+\eta$ and each terms in the model is identified with the corresponding physical force so that the contribution of different forces to interface evolution can be clarified. To reveal the physical pathway through which surfactants regulate the RTI and FI, the reduced model is employed to separate the effects of Marangoni stress $f_{M}$, oscillatory force $f_{O}$, capillary force $f_{C}$, inertial force $f_{I}$, gravity $f_{G}$, and viscous force $f_{V}$ in the nonlinear regime, thereby elucidating the dominant mechanism governing the surfactant induced modification of interfacial instability. Before performing the nonlinear numerical simulations, the following initial conditions are prescribed :

\begin{equation}
\eta(x,0) = -0.1 \cos(kx), \qquad 
q(x,0) = 0, \qquad 
\Gamma(x,0) = 1 ,
\label{eqs5.4}
\end{equation}
where $k$ denotes the nondimensional wavenumber defined under the scaling adopted in Appendix \ref{appA}, and the spatial domain is $0 \le x \le 2\pi/k$. Under periodic boundary conditions, the system is solved by employing a Newton iteration combined with a Fourier spectral method, utilizing 128–256 Fourier modes and a relative tolerance of $10^{-8}$.

\subsection{Rupture dynamics of surfactant-laden films in the absence of oscillation}
% fig8
\begin{figure}
\centerline{\includegraphics[width=0.8\linewidth]{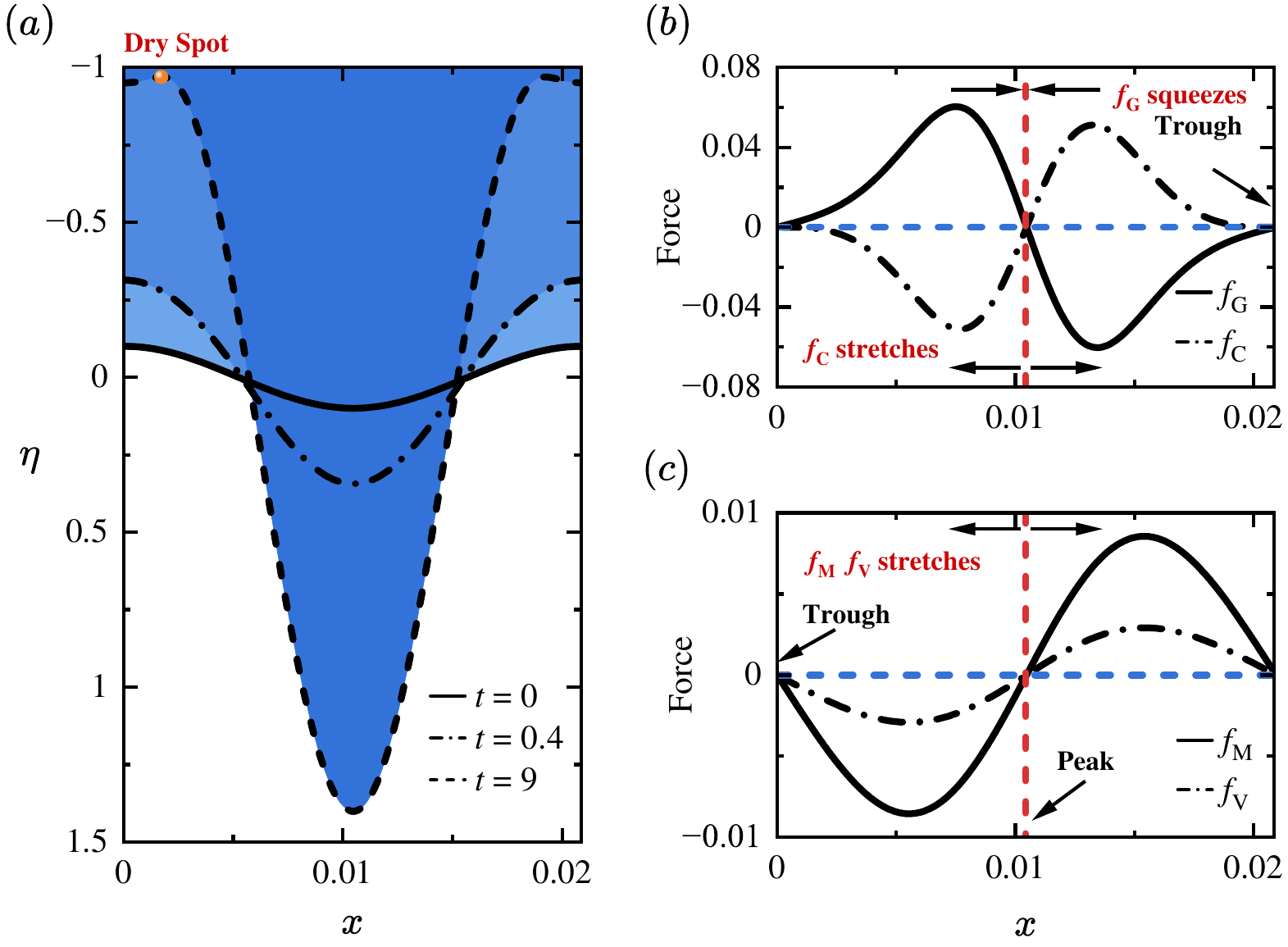}}
  \caption{($a$) Temporal evolution of the liquid film interface, showing the development of a dry spot. 
($b$) Spatial distributions of the gravity $f_G$ and capillary forces $f_C$, at $t=0.4$. 
($c$) Spatial distributions of the Marangoni stress $f_M$ and viscous forces $f_V$, at $t=0.4$.
 }
  \label{fig8}
\end{figure}

Given that the influence of vertical oscillations on the nonlinear evolution of the RTI has been investigated by \cite{sterman2017rayleigh}, this subsection focuses on the role of surfactants in the nonlinear evolution of RTI. It should be noted that the Fourier spectral method is unable to resolve the film dynamics once the minimum film thickness $(1+\eta)_{\min}$ falls below $10^{-2}$. Therefore, the computation is terminated when $(1+\eta)_{\min}=0.05$, which is taken as an approximate rupture time. 

Figure \ref{fig8} presents the spatial distribution of the interfacial disturbance and the corresponding force distributions at $t=0.4$ for $\mathrm{Ma}=50$. 
It is observed from \autoref{fig8}($a$) that fluid near the edges of the film migrates toward the centre, leading to the formation of troughs at the edges and a peak at the centre. The accumulation of fluid at the peak results in significant depletion in the trough region, ultimately giving rise to the formation of a dry spot and the rupture of the liquid film. Further insight is provided by the spatial distributions of the forces at $t=0.4$, shown in \autoref{fig8}($b-c$). The evolution is governed primarily by the combined effects of gravity, capillary force, Marangoni stress, and viscous force, while inertial effects remain negligibly small throughout the process. An examination of the force distributions reveals that gravity is the only destabilizing contribution, driving fluid from the edge troughs toward the central peak and leading to rupture that initiates near the edges. In contrast, capillary force, Marangoni stress, and viscous force act against gravity, transporting fluid from the peak back toward the trough and thereby exerting a stabilizing influence on the film. However, the Marangoni contribution remains relatively weak and is insufficient to fully counteract the destabilizing effect of gravity.

\subsection{Physical mechanism underlying Faraday wave formation}
\label{sec5.2}

As demonstrated in \cite{sterman2017rayleigh}, when vertical oscillations are imposed on a subplate liquid film, the nonlinear evolution within the RTI regime can be classified into three distinct scenarios, depending on the oscillation amplitude. Between the critical amplitudes for RTI and Faraday instability (FI), there exists a stable interval in which the film remains stable in the long-term limit. When the imposed oscillation amplitude is slightly below the critical value for RTI, low wavenumber Faraday waves are excited. When the oscillation amplitude is slightly above the critical value for FI, high wavenumber Faraday waves are generated. In both cases, the film does not rupture but instead undergoes periodic oscillations about a finite mean thickness, and this regime is commonly referred to as the saturated state. Outside this stable interval, the film becomes unstable and eventually ruptures during its long-term evolution.  It is worth emphasizing that the saturated state provides the most suitable framework for analyzing the relative contributions of the various physical forces involved. Therefore, prior to examining the effects of surfactants on interfacial stability under oscillatory forcing, the formation mechanism of Faraday waves must first be clarified.

\begin{figure}
 \centering 
\includegraphics[width=1\textwidth]{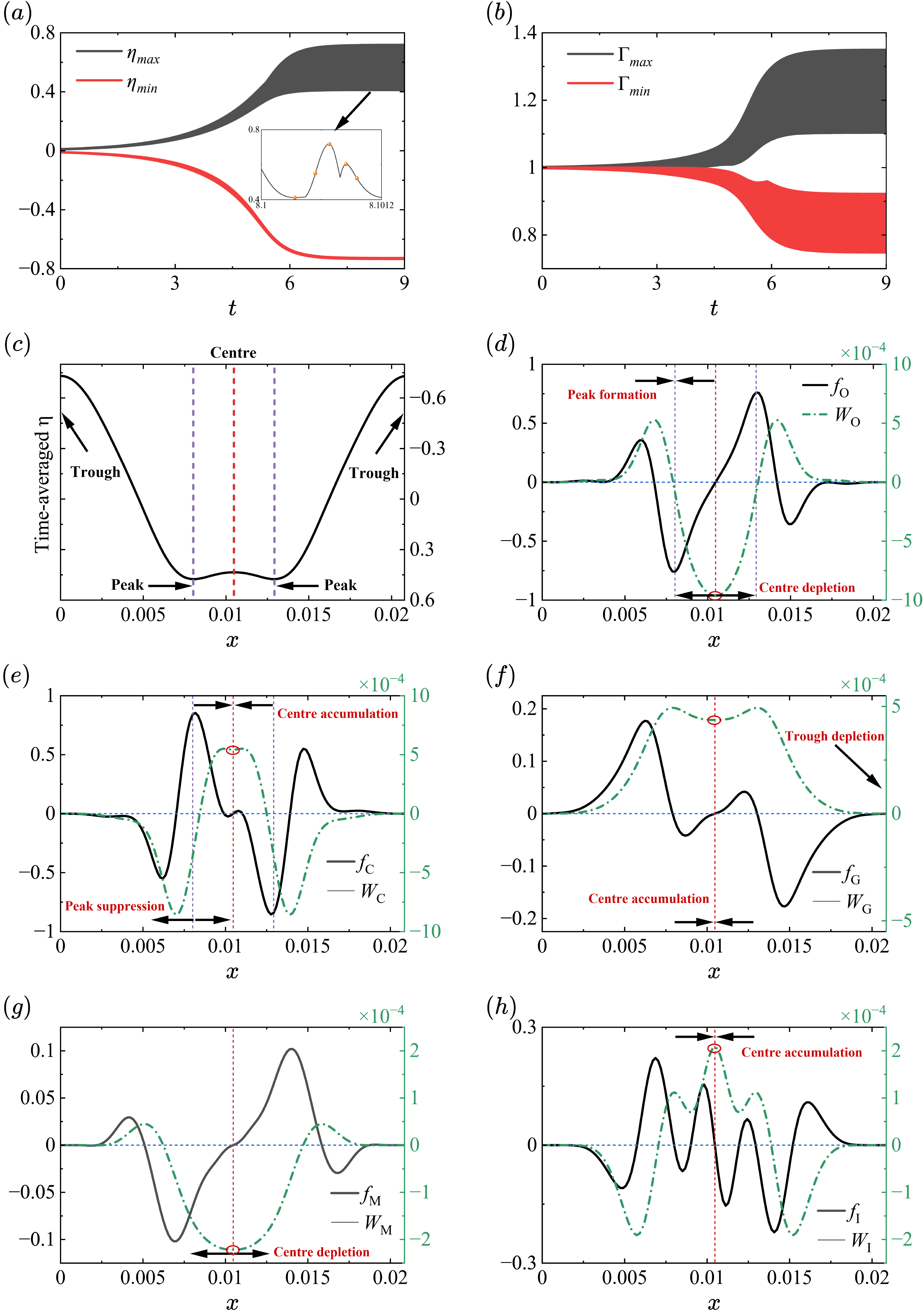}
\caption{ Saturated state results and force analysis in the RTI regime at low frequencies, with $\mathrm{Ma}=10,\ f=10\ \mathrm{Hz},\ a/g=20, k=0.3$.
($a$) Time evolution of the maximum and minimum values of the interface disturbance $\eta$.
($b$) Time evolution of the maximum and minimum values of the surfactant concentration $\Gamma$.
($c$) Spatial distribution of the time averaged $\eta$ in the saturated state.
Individual spatial distributions of the time-averaged forces and their corresponding work:
($d$) oscillatory force $f_O$ and work $W_O$,
($e$) capillary force $f_C$ and work $W_C$,
($f$) gravity $f_G$ and work $W_G$,
($g$) Marangoni force $f_M$ and work $W_M$,
and ($h$) inertial force $f_I$ and work $W_I$.
}
\label{fig9}
\end{figure}

As shown in \autoref{fig9}($a–b$), the maximum values of the free surface $\eta$ and the surfactant concentration $\Gamma$ increase gradually in time, while their minimum values decrease. After sufficiently long evolution, the system reaches a saturated state, in which the maxima of both $\eta$ and $\Gamma$ exhibit periodic temporal variations. 
For an oscillatory system, the time-averaged work performed by the forces is of primary importance. A positive time-averaged contribution corresponds to a net energy input into the system and thus promotes instability, whereas a negative contribution exerts a stabilizing effect. Accordingly, the work done by the time-averaged forces is defined as follows:
\begin{equation}
{W}(x) = \int_{0}^{x} \left( \frac{1}{T} \int_0^T f(\xi,t)\,\mathrm{d}t \right)\mathrm{d}\xi.
\label{eqs5.2.1}
\end{equation}

\autoref{fig9}($c$) shows the time-averaged spatial distribution of the free-surface position $\eta$ in the saturated state. The profile of $\eta$ is symmetric about the centre, with troughs located at the edges of the film and Faraday-wave peaks forming on either side of the centre. The interface exhibits a slight depression at the centre, corresponding to the secondary trough of the excited Faraday wave.

 Further insight into the underlying mechanisms is provided by \autoref{fig9}($d–h$), which shows the time-averaged spatial distributions of the individual forces together with their corresponding work distributions. The time-averaged oscillatory force $f_O$ and the corresponding work $W_O$ are shown in \autoref{fig9}($d$). In the left half of the film centre, $f_O$ is first positive and then negative, indicating that the oscillatory force extracts fluid from both the trough region and the film centre, while exerting a compressive effect on the interface at the force zero-crossing. This process promotes the formation of Faraday-wave peaks. A phase difference is observed between the zero-crossing of $f_O$ and the peak. In contrast, the zero-crossing of $W_O$ is nearly aligned with the peak, indicating that, in a time-averaged sense, the work distribution provides a more direct description of the interface morphology than the force distribution. Specifically, $W_O$ indicates fluid accumulation in the peak region, demonstrating that the oscillatory forcing plays a dominant role in the formation of Faraday waves. In contrast, the capillary force $f_C$ and the corresponding work $W_C$ are shown in \autoref{fig9}($e$). The work associated with $f_C$ exhibits an opposite spatial distribution to that of the $f_O$, leading to the transport of fluid away from the peak toward both the edge troughs and the film centre, which weakens the peak structure. From the perspective of Faraday wave formation, the capillary force suppresses the formation of Faraday-wave peaks.

Note that this behaviour is not inconsistent with the predictions from linear stability analysis, according to which increasing oscillatory force enhances system stability within the RTI regime. The key point is that the promotion of Faraday-wave formation does not necessarily imply an enhancement of system instability. Although oscillatory force facilitates peak formation, the cumulative work at the film centre remains negative, indicating that fluid is preferentially removed from the central region over an oscillation period. Such a central depletion pattern suppresses the growth of overall instability at the early stage of disturbance development. Therefore, oscillatory force does not act as a purely destabilizing mechanism, but instead stabilizes the system at the expense of forming Faraday-wave structures. In other words, the formation of Faraday-wave structures is accompanied by the suppression of system instability, thereby driving the system from a rupture state toward a saturated state. In this sense, the emergence of Faraday waves in the RTI regime represents an intermediate structural state during the transition from instability to stability as the oscillatory force increases. In comparison, the role of capillary force remains consistent, which suppresses the growth of interface peaks in both non-oscillatory and oscillatory configurations, corresponding to the suppression of the primary peak in the former and Faraday-wave peaks in the latter. Capillary force thus acts as a stabilizing mechanism in both cases.

Furthermore, as shown in \autoref{fig9}($f-h$), the work $W_G$ and $W_I$ associated with gravity and inertial force is positive at the film centre, indicating a net transfer of energy toward the central region, which promotes the destabilization of the system. In contrast, the Marangoni force $f_M$ exhibits predominantly negative work at the centre, leading to fluid removal from this region, thus acting as a stabilizing mechanism.
%fig10
\begin{figure}
\centerline{\includegraphics[width=1\linewidth]{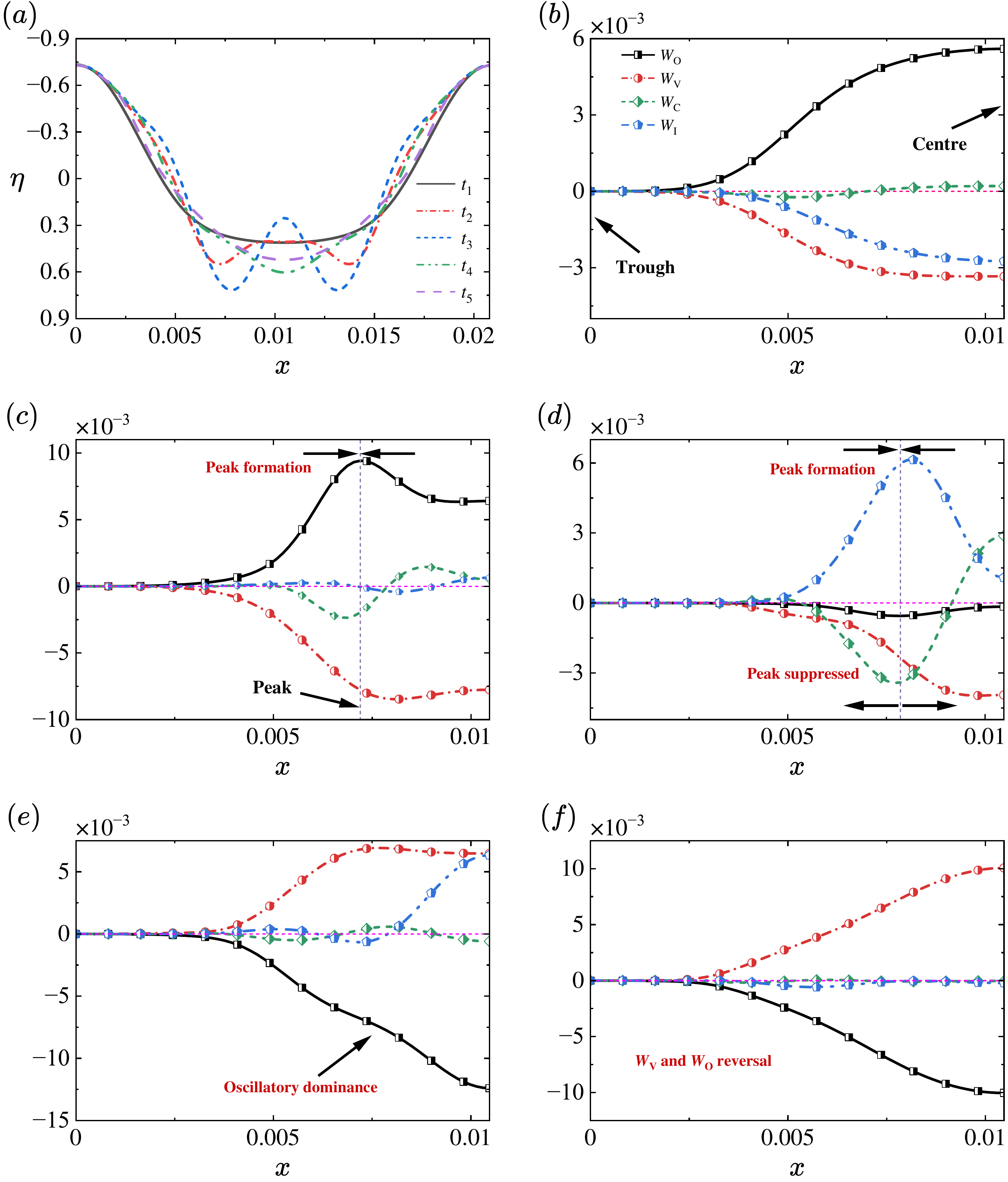}}
  \caption{($a$) Spatial distribution of the interface profile $\eta$ over one oscillation period. Instantaneous spatial distributions of $W_O$, $W_V$, $W_C$, and $W_I$ are shown. Panels ($b$–$d$) correspond to $t_1$–$t_3$ during the growth stage of the disturbance, while ($e$–$f$) correspond to $t_4$–$t_5$ during the decay stage.
 }
  \label{fig10}
\end{figure}

Although previous studies have shown that viscous dissipation increases the critical wavenumber and the critical amplitude of the FI \citep{Kumar1994}, the contribution of the time-averaged viscous force is so weak in the preceding analysis of time-averaged forces and the corresponding work that it can be neglected. To further clarify the role played by viscous effects in the formation of Faraday waves, five representative instants, denoted by $t_1$--$t_5$, are selected within one oscillation period of the saturated state, as shown in \autoref{fig10}($a$). Among these instants, $t_1$--$t_3$ correspond to the growth stage of the Faraday wave, whereas $t_4$--$t_5$ correspond to the decay stage. Since the force distributions are approximately symmetric about the film centre, only half of the spatial period is shown in \autoref{fig10}($b-f$). Four dominant work contributions are displayed, namely the oscillatory work $W_O$, the viscous work $W_V$, the capillary work $W_C$ and the inertial work $W_I$, while the work associated with Marangoni stress and gravity is omitted because of negligible magnitude.

During the initial growth stage of the Faraday wave, as shown in \autoref{fig10}($b-c$), the oscillatory work is dominant and transports fluid from the edge trough towards the film centre, thereby providing the energy required for peak growth. By contrast, viscous work and inertial work oppose this transport. As the evolution proceeds to $t_2$, the influence of capillary force becomes noticeable, while the trend of $W_O$ near the peak undergoes a marked change, reflecting the strengthening of the compressive action exerted by the instantaneous oscillatory force at the peak location. Such a change drives the interface away from the single-peak configuration and towards the double-peak Faraday-wave structure.

Once the Faraday wave has become fully established, as shown in \autoref{fig10}($d$), the oscillatory work is no longer significant and is governed primarily by the combined action of inertial and capillary forces. Inertial effects, which produce a compressive action, favour the formation of Faraday-wave peaks, whereas capillary effects, which act in a stretching manner, suppress such structures. Throughout this stage, viscous work continues to play a dissipative role, opposing the transport of fluid from the edge troughs towards the centre.

During the decay stage of the Faraday wave, as shown in \autoref{fig10}($e-f$), both the oscillatory work and the viscous work undergo reversal. Under such conditions, the oscillatory force transports fluid from the film centre back towards the edge troughs, whereas viscous effects continue to resist this redistribution. Examination of the work distributions over all five instants shows that  oscillatory work during the decay stage is larger than that during the growth stage, while viscous work acts as a dissipative contribution throughout the entire cycle. This observation explains why the time-averaged oscillatory force exhibits an overall stabilizing effect, whereas the time-averaged viscous force becomes negligibly small.

Above all, these results show that the formation of Faraday waves is essentially an energy-accumulation process driven by oscillatory force and realized only after viscous dissipation has been overcome\citep{Kumar1994}. In the initial stage, oscillatory force dominates and supplies the energy required for Faraday-wave growth. Once the Faraday wave has fully developed, the interface morphology is governed mainly by inertial and capillary forces. During the decay stage, the oscillatory force reverses and dominates, with the associated work distribution ultimately determining the stabilizing effect observed in the time-averaged sense.

\subsection{Role of Marangoni effects in the nonlinear dynamics}
%fig11
\begin{figure}
\centerline{\includegraphics[width=0.6\linewidth]{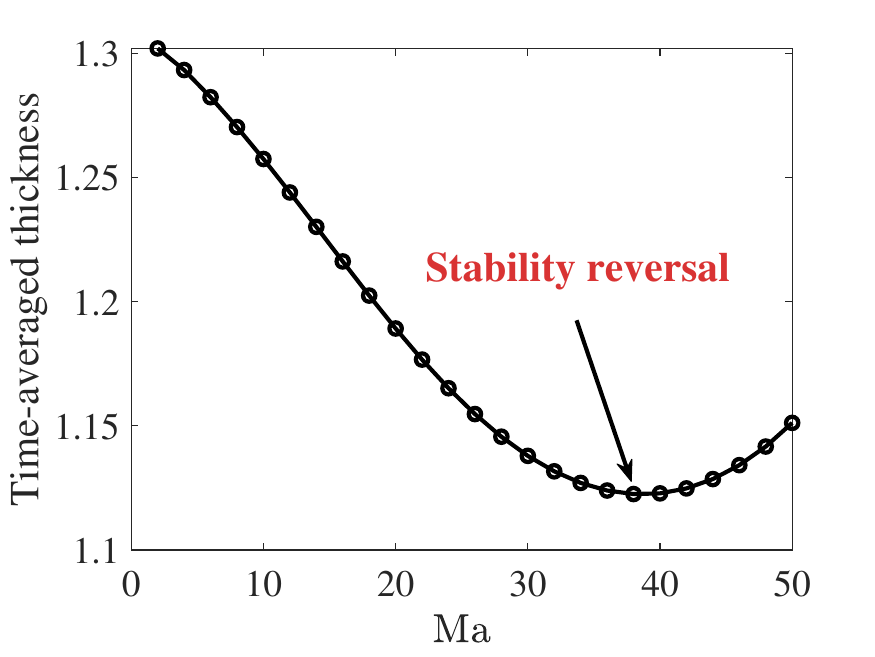}}
  \caption{ Variation of the difference between the maximum and minimum time-averaged $\eta$ with $Ma$, where $a/g = 20$, $f = 10\mathrm{Hz}$ and $k = 0.3$.
 }
  \label{fig11}
\end{figure}

In this subsection, three phenomena are examined that require further clarification. First, the critical amplitude of RTI is found to be nearly independent of $\mathrm{Ma}$. Second, increasing $\mathrm{Ma}$ promotes the onset of FI in the low-frequency regime. Third, in contrast, increasing $\mathrm{Ma}$ suppresses FI in the high-frequency regime.
\begin{figure}
\centerline{\includegraphics[width=1\linewidth]{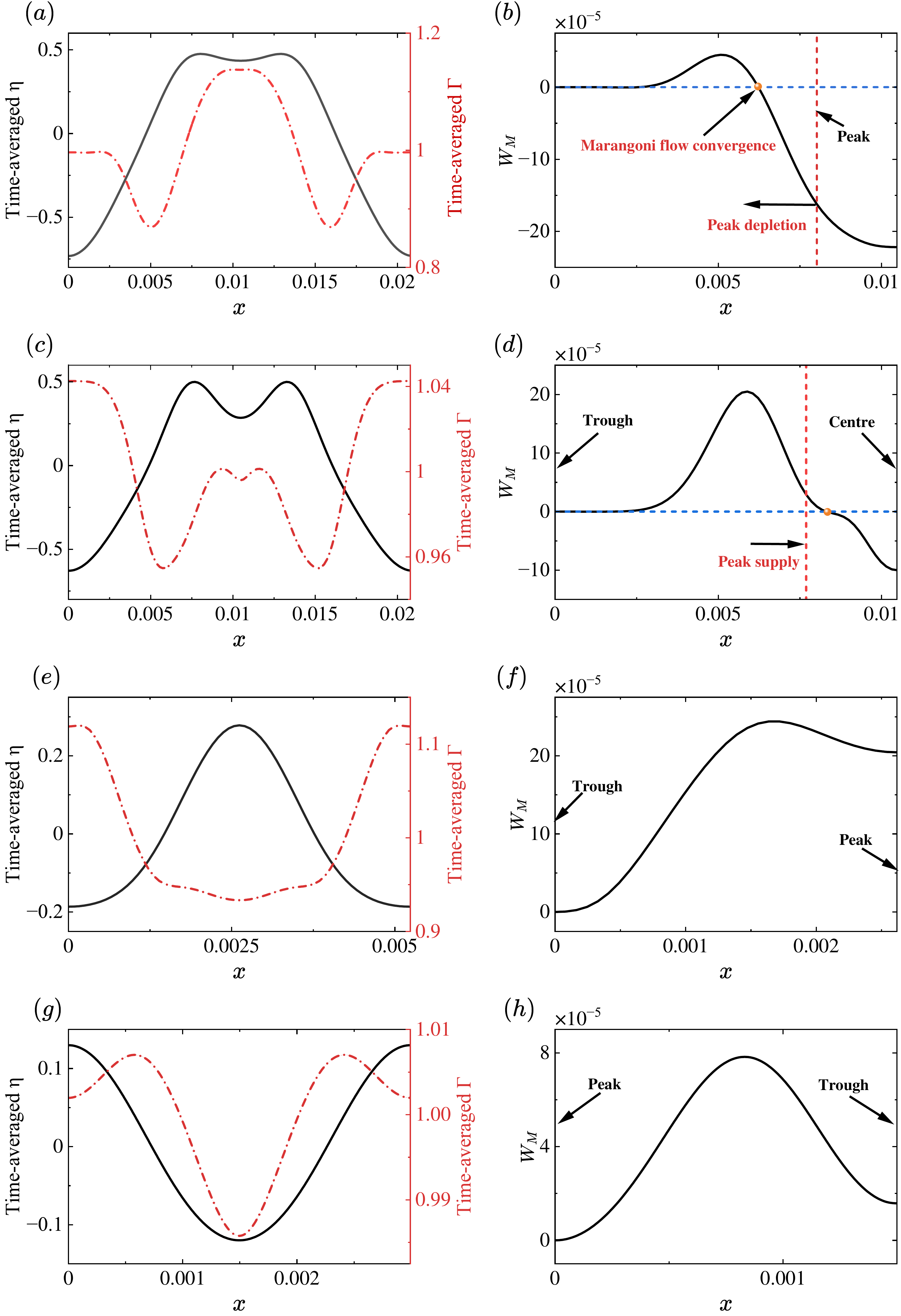}}         
\caption{Spatial distributions of the time averaged $\eta$, the time averaged surfactant concentration $\Gamma$, and the work done by Marangoni stress in the saturated state at a wavenumber of $k = 0.3$. Results are shown for different sets of control parameters: ($a$-$b$) $f = 10\mathrm{Hz}$, $a/g = 20$, $Ma = 10$; ($c$-$d$) $f = 10\mathrm{Hz}$, $a/g = 20$, $Ma = 50$; ($e$-$f$) $f = 10\mathrm{Hz}$, $a/g = 38$, $Ma = 20$; and ($g$-h) $f = 50\mathrm{Hz}$, $a/g = 70$, $Ma = 100$.}
    \label{fig12}
\end{figure}

To address the first point, the time-averaged interface thickness in the saturated state is introduced as a quantitative measure, defined as the difference between the maximum and minimum values of the time-averaged $\eta$. As shown in \autoref{fig11}, this quantity exhibits a non-monotonic dependence on $\mathrm{Ma}$, first decreasing and then increasing. This behaviour indicates that the stabilizing effect at low $\mathrm{Ma}$ remains insufficient to fully suppress the formation of Faraday waves, while at higher $\mathrm{Ma}$ the underlying mechanism undergoes a transition from stabilization to destabilization. This is therefore believed to be the underlying reason for the first phenomenon. 

To elucidate the origin of this stability reversal, \autoref{fig12}($a-d$) presents the spatial distributions of the time-averaged interface position $\eta$, the surfactant concentration $\Gamma$, and the corresponding Marangoni work for $\mathrm{Ma}=10$ and $\mathrm{Ma}=50$. Compared with the oscillatory work shown in \autoref{fig9}($d$), the magnitude of the Marangoni stress work in \autoref{fig12}($b$) is considerably smaller, indicating that Marangoni stress do not play a dominant role in determining the onset of Faraday waves. Consequently, at low $\mathrm{Ma}$, the Marangoni effect provides only limited stabilization. A key difference emerges from the comparison between \autoref{fig12}($a$) and \autoref{fig12}($c$). For $\mathrm{Ma}=10$, the surfactant concentration is higher at the centre and lower near the edges, which remains approximately in phase with the interface deformation. In contrast, for $\mathrm{Ma}=50$, the concentration is lower at the centre and higher near the edges, which becomes out of phase with the interface. This change in phase relation leads directly to a reversal in the spatial distribution of Marangoni work, as shown in \autoref{fig12}($b$) and \autoref{fig12}($d$). Specifically, for $\mathrm{Ma}=10$, the zero of the Marangoni work corresponds to the convergence point of Marangoni-driven flow, located to the left of the Faraday-wave peak. Under this condition, fluid is removed from the peak region, thereby suppressing peak growth. For $\mathrm{Ma}=50$, the convergence point shifts to the right of the peak, resulting in fluid transport towards the peak and consequently promoting the development of Faraday waves. This observation suggests that the stability reversal occurs when the Marangoni convergence point becomes aligned with the peak location, at which point the convergent Marangoni transport most effectively promotes peak growth, thereby providing a physical interpretation of the transition point identified in \autoref{fig11}.

To further investigate the latter two phenomena, the saturated state is triggered by applying a forcing amplitude that exceeds the critical amplitude of FI within the long-wave regime. The mechanisms by which $\mathrm{Ma}$ influences FI under low and high frequency conditions are illustrated in \autoref{fig12}($e-f$) and ($g-h$), respectively. As shown in \autoref{fig12}($e-f$), in the low-frequency regime, the surfactant concentration distribution is nearly out of phase with the interface profile. As a result, the Marangoni stress continuously drives fluid from the troughs of the Faraday wave towards the peaks, thereby amplifying the interfacial instability. In contrast, under high-frequency conditions, the regions of low surfactant concentration are nearly aligned with the lowest points of the interface. The Marangoni-induced flow then transports fluid from the vicinity of the peaks towards the troughs, which reduces interfacial deformation and suppresses the instability of the liquid film, as shown in \autoref{fig12}($g-h$).

\section{Conclusion}\label{Conclusion}

In this study, we have systematically investigated the interfacial dynamics of a Rayleigh–Taylor unstable Newtonian liquid film subjected to the combined effects of vertical oscillatory acceleration and insoluble surfactants. While previous studies have largely treated dynamic stabilization and surfactant-induced Marangoni effects in isolation, the present work bridges this gap, revealing a rich spectrum of coupled modal transitions and novel physical mechanisms through linear Floquet analysis, long-wave asymptotics, and nonlinear simulations.

At the linear level, we identified a profound mode-selective modulation induced by Marangoni stresses. With increasing Marangoni number ($\mathrm{Ma}$), subharmonic Faraday modes are strongly damped, causing the system to transition to a harmonic-dominated regime. Crucially, the topological evolution of the stability boundaries in the amplitude-wavenumber space exhibits a striking frequency dependence. In the low-frequency regime, large Marangoni stresses induce the merging of harmonic instability tongues into a distinct surfactant mode. This mode migrates towards the long-wave regime and eventually coalesces with the Rayleigh-Taylor instability (RTI) branch, which fragments and destroys the parameter window for dynamic stabilization. In contrast, in the high-frequency regime, the addition of surfactants monotonically raises the critical amplitude for the onset of the harmonic instability, effectively broadening the stable window and facilitating the complete suppression of RTI.

To explain the physical origin of surfactant mode migration and coalescence, we derived a long-wave asymptotic model that factorizes the critical instability threshold into a static capillary-gravity margin and a dynamic elasto-inertial modulation. This factorization reveals that the topological transitions are dictated by a critical wavenumber $k_c \propto \omega/\sqrt{\mathrm{Ma}}$, which expresses a fundamental dynamic balance between the destabilizing effective inertia and the stabilizing Marangoni elasticity. As $\mathrm{Ma}$ increases at low frequencies, $k_c$ penetrates the static RTI cutoff, analytically predicting the onset of mode coalescence and the fragmentation of the stable domain.

Beyond the linear and asymptotic regimes, the nonlinear evolution of the interface is explored using a rigorous weighted-residual reduced-order model. By introducing a spatial evaluation of the time-averaged work performed by individual physical forces, the energetic pathways governing the dynamically saturated state are unpacked. It is demonstrated that the formation of Faraday waves within the RTI regime is essentially an energy-accumulation process primarily driven by the time-averaged oscillatory work, which reshapes the interface into multi-peak structures against viscous dissipation and capillary smoothing.

Most importantly, the nonlinear force–work analysis reveals that surfactants modulate interfacial stability through phase-controlled Marangoni transport. In the RTI regime, increasing $\mathrm{Ma}$ induces a phase shift that reverses the direction of transport, and as the associated convergence point becomes aligned with the wave peak, fluid is preferentially driven into the peak region, resulting in a transition from stabilization to destabilization. In the FI regime, the response exhibits a strong frequency dependence, with Marangoni-driven flows removing fluid from the peaks and suppressing the instability at high frequencies, but driving fluid into the peaks and thereby enhancing the instability at low frequencies.

In summary, the present work provides a comprehensive theoretical framework for the control of gravitationally unstable interfaces. The findings emphasize that effective dynamic stabilization of RTI in surfactant-laden systems cannot be achieved by simply maximizing the oscillation amplitude or surfactant concentration. Instead, it requires a delicate tuning of high-frequency forcing to exploit the stabilizing phase dynamics of Marangoni flows. These insights hold significant potential for optimizing interfacial control strategies in applications ranging from advanced propulsion systems to microfluidic material processing.

%\backsection[Acknowledgements]{The authors would like to thank Professor Luca Brandt for his valuable comments and constructive suggestions on this work. His insightful feedback contributed to improving the clarity and rigor of the paper.}

\backsection[Funding]{This research was supported by the National Natural Science Foundation of China (grant nos.12272026, U2341281, 124B2045) and the Natural Science Foundation of Beijing Municipality (grant no. L248008) and the Fundamental Research Funds for the Central Universities. }

\backsection[Declaration of interests]{The authors report no conflict of interest.}

\backsection[Author ORCIDs]{Jun Gao, https://orcid.org/0000-0002-5227-4375; Senlin Zhu, https://orcid.org/0009-0005-4810-2406; 
Luca Brandt, https://orcid.org/0000-0002-4346-4732;
Jianjun Tao, https://orcid.org/0000-0003-4605-1076;
Qingfei Fu, https://orcid.org/0000-0003-2041-2961;
Lijun Yang, https://orcid.org/0000-0002-4625-4028.}

\appendix

\section{Derivation of Reduced Model by Weighted Residual Method }\label{appA}

To derive the reduced model, the following scaling is introduced :
 
\begin{equation}
v = \delta \tilde{v}, \quad 
t =  \tilde{t} / \delta, \quad 
x =  \tilde{x} / \delta, \quad 
\omega = \delta \tilde{\omega} 
\label{eqsA1}
\end{equation}
where $\delta=10^{-4}$ is a small parameter, in a manner similar to \cite{Ignatius2024}. Combining with equation (\ref{eqs3.1}), the following nondimensional governing equations are obtained :

\begin{equation}
u_{x} + v_{y} = 0,
\label{eqsA2}
\end{equation}

\begin{equation}
\delta \mathrm{Re}\left( u_{t} + u u_{x} + v u_{y} \right)
= -\delta p_{x} + \delta^{2} u_{xx} + u_{yy},
\label{eqsA3}
\end{equation}

\begin{equation}
\delta^{2} \mathrm{Re}\left( v_{t} + u v_{x} + v v_{y} \right)
= -p_{y} + \delta^{3} v_{xx} + \delta v_{yy} + B - A \cos(\omega t),
\label{eqsA4}
\end{equation}

\begin{equation}
\Gamma_{t} + u \Gamma_{x} 
+ \frac{\Gamma}{1 + \delta^{2} \eta_{x}^{2}}
\left[ u_{x} + \eta_{x} u_{y} 
+ \delta^{2} \eta_{x} (v_{x} + \eta_{x} v_{y}) \right]
= \frac{\delta}{P e \sqrt{1 + \delta^{2} \eta_{x}^{2}}}
\left( \frac{\Gamma_{x}}{\sqrt{1 + \delta^{2} \eta_{x}^{2}}} \right)_{x}.
\label{eqsA5}
\end{equation}

Then the following boundary conditions are imposed. On the wall at $y=-1$ :

\begin{equation}
u = v = 0 .
\label{eqsA6}
\end{equation}

At the free surface $y=\eta(x,t)$, the dynamic boundary condition and stress balance condition are satisfied :

\begin{equation}
v = \eta_t + u\,\eta_x,
\label{eqsA7}
\end{equation}

\begin{equation}
- p + \frac{2}{1+\delta^2\eta_x^2}
\Big[ \delta^3 \eta_x^2 u_x + \delta v_y - \delta \eta_x (\delta^2 v_x + u_y) \Big]
= \Big[\frac{1}{\mathrm{Ca}} + \mathrm{Ma}(1-\Gamma)\Big]
\frac{\delta^2 \eta_{xx}}{(1+\eta_x^2)^{3/2}},
\label{eqsA8}
\end{equation}

\begin{equation}
2\delta^2 \eta_x (v_y - u_x) + (1 - \delta^2 \eta_x^2)(\delta^2 v_x + u_y)
= -\sqrt{1+\delta^2\eta_x^2}\,\delta\,\mathrm{Ma}\,\Gamma_x.
\label{eqsA9}
\end{equation}

The objective is to derive a reduced model with an accuracy of at least the order of $\delta^2$. Considering that the axial pressure gradient in equation (\ref{eqsA3}) is already of the order of $\delta$, higher-order terms beyond $\delta$ in equation (\ref{eqsA4}) are neglected. Meanwhile, applying boundary conditions of the order of $\delta$, the pressure $p$ can be estimated as :

\begin{equation}
p = {\left. \delta \left( v_{y} - 2 \eta_{x} u_{y} \right) \right|}_{y = \eta}
- \frac{\delta^{2}}{Ca} \eta_{xx}
+ \delta v_{y}
- \left[ B - A \cos(\omega t) \right] (\eta - y)
\label{eqsA10}
\end{equation}

Substituting the expression of $p$ into equation (\ref{eqsA3}) and retaining terms up to the order of $\delta^2$ yields :

\begin{equation}
\begin{aligned}
\delta \operatorname{Re}\left( u_t + u u_x + v u_y \right)
&= -{{\delta }^{2}}\partial_x
   {{\left. \left( {{v}_{y}} - 2{{\eta }_{x}}{{u}_{y}} \right) \right|}_{y=\eta }}
   + \frac{{{\delta }^{3}}}{Ca}{{\eta }_{xxx}}  \\
&\quad + \delta \left[ B - A\cos \left( \omega t \right) \right]{{\eta }_{x}}
   + 2{{\delta }^{2}}u_{xx} + u_{yy}.
\label{eqsA11}
\end{aligned}
\end{equation}

In the above equation, the momentum equation is partially simplified. However, it essentially remains a two-dimensional model, and the velocity components $u$ and $v$ are still unknown. Therefore, in the following part, a reasonable assumption for the velocity distribution $u$ is introduced, and the momentum equation is transformed into a relation between the film thickness $\eta$ and the flow rate $q$ through integration. The velocity $u$ is assumed in the following form:

\begin{equation}
u = \frac{3q}{\eta + 1} \left[ \frac{y+1}{\eta + 1} - \frac{1}{2} \left( \frac{y+1}{\eta + 1} \right)^2 \right]
+ \frac{\delta \mathrm{Ma} \, \Gamma_x (\eta + 1)}{2} \left[ \frac{y+1}{\eta + 1} - \frac{3}{2} \left( \frac{y+1}{\eta + 1} \right)^2 \right],
\label{eqsA12}
\end{equation}
where $q=\int_{-1}^{\eta }{udy}$. Then, the velocity $v$ can be obtained from the continuity equation :

\begin{equation}
\begin{aligned}
v &= -\int_{-1}^{\eta} u_x \, dy = -\Bigg\{ 
      \frac{(2 + 3\eta - y)(1+y)^2}{2(1+\eta)^3} q_x
      - \frac{3(1 + 2\eta - y)(1+y)^2}{2(1+\eta)^4} q \, \eta_x \\
  &\quad + \delta \mathrm{Ma} \, \Gamma_{xx} \frac{(\eta - y)(1+y)^2}{4(1+\eta)}
      + \delta \mathrm{Ma} \, \Gamma_x \, \eta_x \frac{(1+y)^3}{4(1+\eta)^2}
    \Bigg\}.
\end{aligned}
\label{eqsA13}
\end{equation}

Next, the reduced model is derived using the weighted residual method, with the first step being the specification of the weighting function. Noting that in equation (\ref{eqsA11}) only $u_{yy}$ is of $O(1)$ magnitude and constitutes the primary source of error in the weighted residual method, the following integral result is obtained via integration by parts :

\begin{equation}
\int_{-1}^{\eta} u_{yy} \, F \, dy = \int_{-1}^{\eta} u \, F_{yy} \, dy + \left[ u_y F - u F_y \right]_{y=-1}^{y=\eta},
\label{eqsA14}
\end{equation}

Therefore, the weighting function is required to satisfy the following condition, which ensures that the order of error in this integral is controlled by the boundary conditions :

\begin{equation}
F_{yy} = 1, \quad F|_{y=-1} = 0, \quad F_y|_{y=\eta} = 0.
\label{eqsA15}
\end{equation}

Thus, the test function is given by:

\begin{equation}
F=\frac{1}{2}\left( {{y}^{2}}-2\eta y-2\eta -1 \right).
\label{eqsA16}
\end{equation}

To ensure that the error order of the weighted integral of $u_{yy}$ is $O(\delta^2)$, the following boundary conditions are imposed:

\begin{equation}
{{\left. u \right|}_{y=-1}}=0,
{{\left. {{u}_{y}} \right|}_{y=\eta }}=-\delta \mathrm{Ma}{{\Gamma }_{x}}-{{\delta }^{2}}\left( {{v}_{x}}+4{{\eta }_{x}}{{v}_{y}} \right)
\label{eqsA17}
\end{equation}

The integral form of equation (\ref{eqsA11}) is then given as:

\begin{equation}
\begin{aligned}
  & \int_{-1}^{\eta} \delta \operatorname{Re} \left( u_t + u u_x + v u_y \right) F \, dy 
    = \int_{-1}^{\eta} \left( \frac{\delta^3}{Ca} \eta_{xxx} + \delta \left[ B - A \cos(\omega t) \right] \eta_x \right) F \, dy \\
  & \quad + \int_{-1}^{\eta} u_{yy} F \, dy 
    + \int_{-1}^{\eta} \left( -\delta^2 \left. \left( v_y - 2 \eta_x u_y \right)_x \right|_{y=\eta} + 2 \delta^2 u_{xx} \right) F \, dy.
\end{aligned}
\label{eqsA18}
\end{equation}

The final result of the integration is given as follows:

\begin{equation}
\begin{aligned}
 & -\delta \mathrm{Re}\Big\{ \frac{2}{5} b^2 q_{t} + \frac{34 b}{35} q q_{x} - \frac{18}{35} q^2 \eta_{x} \Big\} -\delta^3 \mathrm{Re} \mathrm{Ma}^2 \frac{b^4}{336} \Gamma_{x} \Big( \frac{5}{2} b \Gamma_{xx} + \frac{19}{5} \eta_{x} \Gamma_{x} \Big) \\
 & +\delta^2 \mathrm{Re} \mathrm{Ma} \Big\{ \frac{b^3}{120} \big[ b \Gamma_{xt} + \frac{19}{7} q_{x} \Gamma_{x} \big] 
   + \frac{b^2}{56} q \big[ \Gamma_{x} \eta_{x} + \frac{3 b}{2} \Gamma_{xx} \big] \Big\} \\
 & = -\frac{b^3}{3} \Big( \delta B \eta_{x} - \delta A \cos(\omega t) \eta_{x} + \frac{\delta^3}{Ca} \eta_{xxx} \Big) \\
 & \quad + q + \delta^2 \Big\{ \frac{9}{5} b (q_{x} \eta_{x} - b q_{xx}) + \frac{4}{5} q (3 b \eta_{xx} - 2 \eta_{x}^2) \Big\} \\
 & \quad + \frac{1}{2} \delta b^2 \mathrm{Ma} \Gamma_{x} 
   + \delta^3 \mathrm{Ma} \frac{b^2}{5} \Big\{ \frac{b}{2} ( b \Gamma_{xxx} + 3 \eta_{x} \Gamma_{xx} ) + \frac{b}{3} \eta_{xx} \Gamma_{x} - 4 \eta_{x}^2 \Gamma_{x} \Big\},
\end{aligned}
\label{eqsA19}
\end{equation}
where $b=1+\eta$. The continuity equation is then rewritten using the kinematic boundary condition, and the surfactant transport equation is given as follows :

\begin{equation}
{{\eta }_{t}}+{{q}_{x}}=0,
\label{eqsA20}
\end{equation}

\begin{equation}
{{\Gamma }_{t}} + \partial_x\left[ \frac{3 q \Gamma}{2 b} - \frac{\mathrm{Ma}}{4} b \Gamma \Gamma_{x} \right] = \frac{1}{Pe} \Gamma_{xx}.
\label{eqsA21}
\end{equation}
Thus, equations (\ref{eqsA19}–\ref{eqsA21}) constitute the reduced model.

\section{Reliability verification of the reduced model}\label{appB}
% fig13
\begin{figure}
  \centerline{\includegraphics[width=0.6\linewidth]{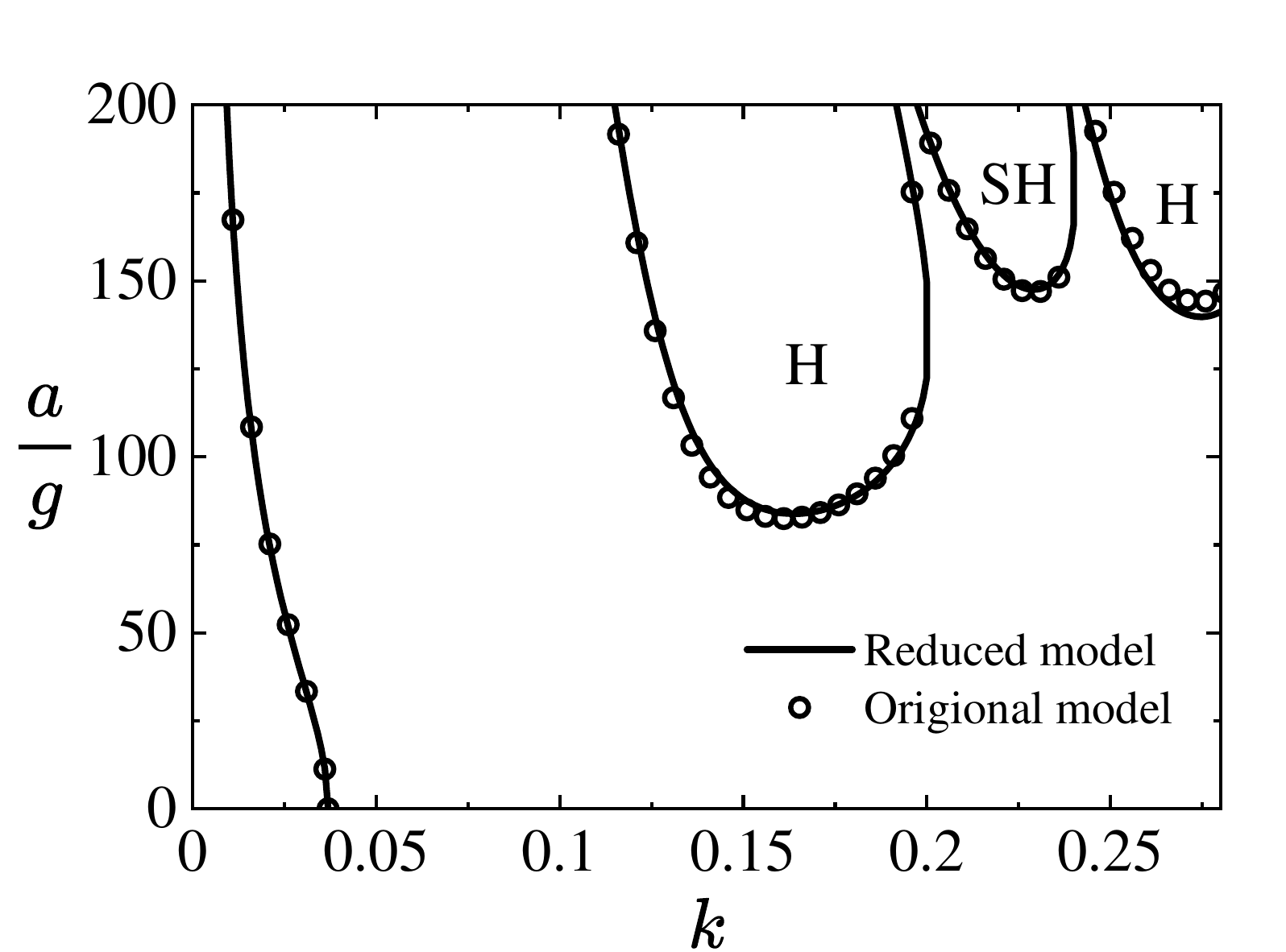}}
  \caption{\raggedright Comparison of neutral stability curves for harmonic (H) and subharmonic (SH) modes between the reduced and original models, with the corresponding parameter values given as follows: $f=50\mathrm{Hz}$, $\mathrm{Ma}=340$.
 }
  \label{fig13}
\end{figure}
To facilitate direct comparison with the linear numerical results of the original model, the variables in the reduced model are first rescaled back to their original dimensional form, after which the following perturbation expansion is introduced: 

\begin{equation}
q(x,y,t) =0 + e^{ikx} \sum_{n=-\infty}^{\infty} q_n \, e^{(s + i\alpha + i n \omega) t},
\label{eqsB1}
\end{equation}

\begin{equation}
\eta(x,t) =0+ e^{ikx} \sum_{n=-\infty}^{\infty} \eta_n \, e^{(s + i\alpha + i n \omega) t},
\label{eqsB2}
\end{equation}

\begin{equation}
\Gamma(x,t) = 1+e^{ikx} \sum_{n=-\infty}^{\infty} \Gamma_n \, e^{(s + i\alpha + i n \omega) t}.
\label{eqsB3}
\end{equation}

The reduced model is then linearized and simplified, yielding:

\begin{equation}
\sum\limits_{n=-\infty}^{\infty}{{{S}_{n}}{{{\eta }}_{n}}}=\sum\limits_{n=-\infty}^{\infty}{A\left( {{{\eta }}_{n-1}}+{{{\eta }}_{n+1}} \right)},
\label{eqsB4}
\end{equation}
where $S_n$ is given as follows:

\begin{equation}
{{S}_{n}} = \frac{12\mathrm{Re}}{5{{k}^{2}}}
\left\{
\begin{aligned}
  & -\gamma _{n}^{2}
  -\left( \frac{5}{2\mathrm{Re}}{{\gamma }_{n}}
  +\frac{9}{2\mathrm{Re}}{{k}^{2}}{{\gamma }_{n}} \right)
  +\frac{5}{6\mathrm{Re}}{{k}^{2}}\left( B-\frac{1}{Ca}{{k}^{2}} \right) \\[6pt]
  & +{{k}^{2}}\mathrm{Ma}\left( \frac{1}{48}{{\gamma }_{n}}
  -\frac{5}{4\mathrm{Re}}
  +\frac{1}{4\mathrm{Re}}{{k}^{2}} \right)
  \frac{3{{\gamma }_{n}}}{2\left( {{\gamma }_{n}}
  +\frac{1}{Pe}{{k}^{2}}
  +\frac{1}{4}\mathrm{Ma}{{k}^{2}} \right)}
\end{aligned}
\right\}.
\label{eqsB5}
\end{equation}

Subsequently, the matrices are assembled according to equation (\ref{eqs3.5}), and the resulting linear problem is solved. \autoref{fig13} presents a comparison between the numerical solutions of the reduced model and the original model, and the good agreement between the two demonstrates the reliability of the reduced model.

\bibliographystyle{jfm}
\bibliography{jfm}

@Article{LiS2023,
 author      = {Li, S. and Chen, Y.Z. and Cheng, Z. and Peng, J.},
 title       = {The role of soluble surfactant in the linear instability of a film coating inside a tube},
 journal     = {Journal of Fluid Mechanics},
 year        = {2023},
 volume      = {973},
 pages       = {A46}
 }

@article{Sharp1984,
	author = {Sharp, D. H.},
	title = {An overview of {Rayleigh-Taylor} instability},
	journal = {Physica D: Nonlinear Phenomena},
	volume = {12},
	number = {1-3},
	pages = {3--18},
	year = {1984}
}

@article{Cabot2006,
	author = {Cabot, W. H. and Cook, A. W.},
	title = {Reynolds number effects on {Rayleigh–Taylor} instability with possible implications for type {Ia} supernovae},
	journal = {Nature Physics},
	volume = {2},
	number = {8},
	pages = {562--568},
	year = {2006}
}

@article{Betti2016,
	author = {Betti, R. and Hurricane, O. A.},
	title = {Inertial-confinement fusion with lasers},
	journal = {Nature Physics},
	volume = {12},
	number = {5},
	pages = {435--448},
	year = {2016}
}

@article{Zhou2025,
	author = {Zhou, Y. and Sadler, J. D. and Hurricane, O. A.},
	title = {Instabilities and Mixing in Inertial Confinement Fusion},
	journal = {Annual Review of Fluid Mechanics},
	volume = {57},
	pages = {197--225},
	year = {2025}
}

@article{Qin2018,
	author = {Qin, L. Z. and Yi, R. and Yang, L. J.},
	title = {Theoretical breakup model in the planar liquid sheets exposed to high-speed gas and droplet size prediction},
	journal = {International Journal of Multiphase Flow},
	volume = {98},
	pages = {158--167},
	year = {2018}
}

@article{Yang2020,
	author = {Yang, L. J. and Gao, Y. P. and Li, J. X. and Fu, Q. F.},
	title = {Theoretical atomization model of a coaxial gas-liquid jet},
	journal = {Physics of Fluids},
	volume = {32},
	number = {12},
	pages = {124108},
	year = {2020}
}

@article{Hu2024,
	author = {Hu, K. X. and Du, K.},
	title = {Nonlinear waves in a sheared liquid film on a horizontal plane at small {Reynolds} numbers},
	journal = {Journal of Fluid Mechanics},
	volume = {999},
	pages = {A14},
	year = {2024}
}

@article{Zhu2025b,
	author = {Zhu, S. L. and Tao, J. J.},
	title = {{Nonlinear Rayleigh-Taylor instability of pendent finite-thickness film subjected to vibration}},
	journal = {Physics of Fluids},
	volume = {37},
	number = {9},
	pages = {092102},
	year = {2025}
}

@article{Wolf1969,
	author = {Wolf, G. H.},
	title = {The dynamic stabilization of the {Rayleigh-Taylor} instability and the corresponding dynamic equilibrium},
	journal = {Zeitschrift f{\"u}r Physik},
	volume = {227},
	number = {4},
	pages = {291--300},
	year = {1969}
}

@article{sterman2017rayleigh,
	title={Rayleigh-Taylor instability in thin liquid films subjected to harmonic vibration},
	author={Sterman-Cohen, Elad and Bestehorn, Michael and Oron, Alexander},
	journal={Physics of Fluids},
	volume={29},
	number={5},
	year={2017},
	publisher={AIP Publishing}
}

@article{Wolf1970,
	author = {Wolf, G. H.},
	title = {Dynamic stabilization of the interchange instability of a liquid-gas interface},
	journal = {Physical Review Letters},
	volume = {24},
	number = {9},
	pages = {444--446},
	year = {1970}
}

@article{Troyon1971,
	author = {Troyon, F. and Gruber, R.},
	title = {Theory of the dynamic stabilization of the {Rayleigh-Taylor} instability},
	journal = {Physics of Fluids},
	volume = {14},
	number = {10},
	pages = {2069--2073},
	year = {1971}
}

@article{Apffel2020,
	author = {Apffel, B. and Novkoski, F. and Eddi, A. and Fort, E.},
	title = {Floating under a levitating liquid},
	journal = {Nature},
	volume = {585},
	number = {7823},
	pages = {48--52},
	year = {2020}
}

@article{Zhu2024,
	author = {Zhu, S. L. and Tao, J. J.},
	title = {Vibration effect on {Rayleigh–Taylor} instability of sedimenting suspension},
	journal = {Physics of Fluids},
	volume = {36},
	number = {3},
	pages = {033310},
	year = {2024}
}

@article{Zhu2025,
	author = {Zhu, S. L. and Tao, J. J. and Zeng, R. H.},
	title = {{Rayleigh–Taylor} instability subjected to dual-frequency normal vibration},
	journal = {Physics of Fluids},
	volume = {37},
	number = {1},
	pages = {014113},
	year = {2025}
}

@article{Zhu2026,
	title = {{Interface coupling effects on Rayleigh–Taylor and Faraday instabilities in granular suspensions under oscillatory forcing}},
	journal = {International Journal of Multiphase Flow},
	volume = {194},
	pages = {105448},
	year = {2026},
	issn = {0301-9322},
	author = {Zhu, S. L. and Fu, Q. F. and Yang, L. J.},
}

@article{Zhao2024,
	author = {Zhao, K. G. and Li, Z. Y. and Wang, L. F. and Di, Z. H. and Xue, C. and Zhang, H. and Wu, J. F. and Ye, W. H. and Zhou, C. T. and Ding, Y. K. and Zhang, W. Y. and He, X. T.},
	title = {Dynamic stabilization of ablative {Rayleigh-Taylor} instability in the presence of a temporally modulated laser pulse},
	journal = {Physical Review E},
	volume = {109},
	number = {2},
	pages = {025213},
	year = {2024}
}

@article{Zhao2025,
	title = {Dynamic stabilization and parametric excitation of instabilities in an ablation front by a temporally modulated laser pulse},
	author = {Zhao, K. G. and Li, J. W. and Wang, L. F. and Li, Z. Y. and Di, Z. H. and Xue, C. and Dong, J. Q. and Zhang, H. and Wu, J. F. and Zhuo, H. B. and Ye, W. H. and Zhou, C. T. and Ding, Y. K. and Zhang, W. Y. and He, X. T.},
	journal = {Physical Review E},
	volume = {111},
	issue = {5},
	pages = {055204},
	year = {2025},
}

@article{Faraday1831,
	author = {Faraday, M.},
	title = {On a peculiar class of acoustical figures; and on certain forms assumed by groups of particles upon vibrating elastic surfaces},
	journal = {Philosophical Transactions of the Royal Society of London},
	volume = {121},
	pages = {299--340},
	year = {1831}
}

@article{Kumar1994,
	author = {Kumar, K. and Tuckerman, L. S.},
	title = {Parametric instability of the interface between two fluids},
	journal = {Journal of Fluid Mechanics},
	volume = {279},
	pages = {49--68},
	year = {1994}
}

@article{Kumar1996,
	author = {Kumar, K.},
	title = {Linear theory of Faraday instability in viscous liquids},
	journal = {Proceedings of the Royal Society of London. Series A: Mathematical, Physical and Engineering Sciences},
	volume = {452},
	pages = {1113--1126},
	year = {1996}
}

@article{Kudrolli1998,
	title = {Superlattice patterns in surface waves},
	journal = {Physica D: Nonlinear Phenomena},
	author = {A. Kudrolli and B. Pier and J.P. Gollub},
	volume = {123},
	number = {1},
	pages = {99-111},
	year = {1998},
	issn = {0167-2789},
}

@article{Urra2019,
	title = {Localized Faraday patterns under heterogeneous parametric excitation},
	author = {Urra, H\'ector and Mar\'{\i}n, Juan F. and P\'aez-Silva, Milena and Taki, Majid and Coulibaly, Saliya and Gordillo, Leonardo and Garc\'{\i}a-\~Nustes, M\'onica A.},
	journal = {Phys. Rev. E},
	volume = {99},
	issue = {3},
	pages = {033115},
	numpages = {9},
	year = {2019},
}

@article{Ignatius2025, 
	title={Gravitational effects on Faraday instability in a viscoelastic liquid}, 
	volume={1011}, 
	DOI={10.1017/jfm.2025.328}, 
	journal={Journal of Fluid Mechanics}, 
	author={Ignatius, I.B. and Dinesh, B. and Dietze, G.F. and Narayanan, R.}, 
	year={2025}, 
	pages={A28}
}

@article{Lohse2023,
	author = {Lohse, D.},
	title = {Surfactants on troubled waters},
	journal = {Journal of Fluid Mechanics},
	volume = {976},
	pages = {F1},
	year = {2023}
}

@article{chakraborty2020, 
	title={Stability of nanofluid: A review}, 
	author={Chakraborty, Samarshi and Panigrahi, Pradipta Kumar}, 
	journal={Applied Thermal Engineering}, 
	volume={174}, 
	pages={115259}, 
	year={2020}
}

@article{Li2023,
	author = {Li, F. and He, D. D.},
	title = {Dynamics of a surfactant-laden viscoelastic thread in the presence of surface viscosity},
	journal = {Journal of Fluid Mechanics},
	volume = {966},
	pages = {A35},
	year = {2023}
}

@article{Stewart2013,
	author = {Stewart, P. S. and Davis, S. H. and Hilgenfeldt, S.},
	title = {Viscous {Rayleigh-Taylor} instability in aqueous foams},
	journal = {Colloids and Surfaces A: Physicochemical and Engineering Aspects},
	volume = {436},
	pages = {898--905},
	year = {2013}
}

@article{Kumar2002,
	author = {Kumar, S. and Matar, O. K.},
	title = {Parametrically driven surface waves in surfactant-covered liquids},
	journal = {Proceedings of the Royal Society of London. Series A: Mathematical, Physical and Engineering Sciences},
	volume = {458},
	number = {2027},
	pages = {2815--2828},
	year = {2002}
}

@article{Kumar2004,
	author = {Kumar, S. and Matar, O. K.},
	title = {On the {Faraday} instability in a surfactant-covered liquid},
	journal = {Physics of Fluids},
	volume = {16},
	number = {1},
	pages = {39--46},
	year = {2004}
}

@article{Suman2008,
	author = {Suman, B. and Kumar, S.},
	title = {Surfactant- and elasticity-induced inertialess instabilities in vertically vibrated liquids},
	journal = {Journal of Fluid Mechanics},
	volume = {610},
	pages = {407--423},
	year = {2008}
}

@article{Panda2025,
	author = {Panda, D. and Kahouadji, L. and Tuckerman, L. S. and Shin, S. and Chergui, J. and Juric, D. and Matar, O. K.},
	title = {Marangoni-driven patterns, ridges and hills in surfactant-covered parametric surface waves},
	journal = {Journal of Fluid Mechanics},
	volume = {1008},
	pages = {R4},
	year = {2025}
}

@article{Gao2006,
	author = {Gao, P. and Lu, X.-Y.},
	title = {Effect of surfactants on the long-wave stability of oscillatory film flow},
	journal = {Journal of Fluid Mechanics},
	volume = {562},
	pages = {345--354},
	year = {2006}
}

@article{Wang2023,
	author = {Wang, S. and Du, S. and Xiao, Y. and Zhao, M.},
	title = {Instability of a viscoelastic film with insoluble surfactants on an oscillating plane},
	journal = {Journal of Fluid Mechanics},
	volume = {973},
	pages = {A39},
	year = {2023}
}

@Article{Ignatius2024,
 author = {I.B. Ignatius and B. Dinesh, G.F. Dietze and R. Narayanan},
 title = { Influence of parametric forcing on Marangoni
 instability},
 journal     = {J. Fluid. Mech.},
 year        = {2024},
 volume      = {981},
 pages       = {A8}
 }
%\bibliography{jfm2esam}

%% End of file `jfm2esam.bib'.

\end{document}